\documentclass[aps,prb,twocolumn,floatfix,superscriptaddress,showpacs]{revtex4}
\pdfoutput=1
\usepackage{natbib}
\usepackage{colordvi}
\usepackage{graphicx}
\usepackage{epstopdf}
\usepackage{bm}
\usepackage{amsmath, amsfonts}
\usepackage{soul}
\usepackage[colorlinks=true,citecolor=blue,linkcolor=blue]{hyperref}

\begin{document}

\title{$\mathsf{Z}_2$-slave-spin theory for strongly correlated fermions}
\author{A. R\"uegg}
\affiliation{Theoretische Physik, ETH Z\"urich, CH-8093 Z\"urich, Switzerland}
\affiliation{Department of Physics, The University of Texas at Austin, Austin, Texas 78712, USA}
\author{S.D.~Huber}
\affiliation{Theoretische Physik, ETH Z\"urich, CH-8093 Z\"urich, Switzerland}
\affiliation{Department of Condensed Matter Physics, The Weizmann Institute of Science, 76100 Rehovot, Israel}
\author{M. Sigrist}
\affiliation{Theoretische Physik, ETH Z\"urich, CH-8093 Z\"urich, Switzerland}
\pacs{71.27.+a, 71.10.--w, 71.30.+h}
\begin{abstract}
We review a representation of Hubbard-like models that is based on auxiliary pseudospin variables. These pseudospins refer to the local charge {\it modulo two} in the original model and display a local $\mathsf{Z}_2$ gauge freedom. We discuss the associated mean-field theory in a variety of different contexts which are related to the problem of the interaction-driven metal-insulator transition at half-filling including Fermi surface deformation and spectral features beyond the local approximation. Notably, on the mean-field level, the Hubbard bands are derived from the excitations of an Ising model in a transverse field and the quantum critical point of this model is identified with the Brinkman-Rice criticality of the almost localized Fermi liquid state. Non-local correlations are included using a cluster mean-field approximation and the Schwinger boson theory for the auxiliary quantum Ising model.
\end{abstract}
\maketitle
\section{Introduction}
Strong correlation physics is a central ingredient for diverse solid-state systems including the high-$T_c$ cuprates,\cite{Lee:2006b} other transition metal oxides\cite{Imada:1998} or unconventional superconductors\cite{Sigrist:1991} as well as fractional quantum Hall states.\cite{Nayak:2008} Related physics is currently also discussed in the context of ultracold atoms in optical lattices.\cite{Bloch:2008} Strong interactions can give rise to a variety of unusual quantum phases including ordered phases in spin, charge, and orbital degrees of freedom, as well as miscellaneous exotic liquid phases.\cite{Tokura:2000} The complexity of these many-body systems lies in the fundamental dichotomy between real space (localization) and momentum space (delocalization) in combination with the restrictions on the Hilbert space which are enforced by the strong interaction among the particles. Only a few rigorous results are available for comparable ``simple" models such as the Hubbard or the $t$-$J$ Hamiltonian in more than one\cite{Lieb:1968,Giamarchi:2004} or less than infinite dimensions.\cite{Metzner:1989,Georges:1996} Hence, there is a considerable amount of ongoing work attempting to reach the physically most relevant limit of two or three dimensions.\cite{Maier:2005} Thereby, the classical problem of the Mott metal-insulator transition\cite{Hubbard:1963} plays a key role and its nature in various systems is still actively debated. Recent examples include the orbital-selective Mott transition in multiorbital systems such as Ca$_{2-x}$Sr$_x$RuO$_4$,\cite{Nakatsuji:2000} the problem of momentum space differentiation related to the pseudogap phenomena  in cuprates\cite{Damascelli:2003} or the paramagnetic metal-insulator (spin liquid) transition in frustrated geometries such as in the organic compound $\kappa$-(BEDT-TTF)$_2$Cu$_2$(CN)$_3$.\cite{Shimizu:2003}

Widely used theoretical approaches to tackle these problems are slave-particle methods because they provide powerful tools to deal with the restrictions imposed on the Hilbert space due to strong correlations. They have been pioneered in the context of quantum magnets,\cite{Holstein:1940, Auerbach:1994} magnetic impurities in metals\cite{Barnes:1976,Coleman:1984,Read:1983} and doped Mott insulators.\cite{Lee:1992,Senthil:2000} The basic idea is to represent local degrees of freedom with the help of auxiliary degrees of freedom in an enlarged Hilbert space. However, in order to have a faithful representation, these auxiliary degrees of freedom obey certain constraints and are not independent (although in some cases they survive as ``real" particles)\cite{Lee:2006b} but are ``enslaved" -- hence, the name. In practice, the starting point is usually a mean-field state in the enlarged Hilbert space obeying a set of self-consistency equations. This allows for a semi-analytical and non-perturbative treatment of correlation effects.

In this article, we make contact with a particular class of slave-particle representations for Hubbard-like models. They share a simple physical picture of the interaction-driven paramagnetic metal-insulator transition which dates back to early works of Gutzwiller\cite{Gutzwiller:1963} and Brinkman and Rice\cite{Brinkman:1970} and others\cite{Vollhardt:1984} who introduced the notion of an ``almost localized Fermi liquid" to characterize the metallic state close to the Mott transition. In particular, the transition to the localized (insulating) state is signaled by a diverging effective mass. In infinite dimensions, this picture can be put on a more firm ground\cite{Georges:1996} since correlations are strictly local but in two or three dimensions non-local correlations can lead to different conclusions. Our approach is closely related to the four-boson method introduced by Kotliar and Ruckenstein (KR),\cite{Kotliar:1986} and its extensions.\cite{Li:1989,Fresard:1992,Fresard:1997uq,Hasegawa:1997, Rueegg:2005, Lechermann:2007, Rueegg:2007, Rueegg:2008a} In these approaches it is, however, not straightforward to include fluctuations of the mean fields and the high-energy (incoherent) part of the single-particle spectrum. Most studies are therefore restricted to the low-energy (coherent) part of the one-particle spectrum (but note also Refs.~[\onlinecite{Lavagna:1990, Jolicoeur:1991, Castellani:1992, Raimondi:1993, Arrigoni:1995}]). An elegant formulation which can overcome some of these shortcomings has been given by Florens and Georges\cite{Florens:2002, Florens:2004} in terms of a slave-rotor representation. In this formulation, the phase fluctuations of the rotors give rise to the incoherent spectral features in the single-particle spectrum. Moreover, it also yields a closer connection to the superfluid to insulator transition in the Bose-Hubbard model.\cite{Fisher:1989} We also note the slave-spin representation of de'Medici and coworkers \cite{deMedici:2005} which has been introduced to study the orbital-selective Mott transition in a two-band Hubbard model and follows a similar spirit.

We contribute to these different approaches by reviewing an alternative slave-spin formulation which has recently been applied to the study of dynamically generated double occupancy in cold atomic Fermi systems.\cite{Huber:2009} The advantage of our formulation is that it reduces the complexity of the representation to a minimum. The gauge freedom is only $\mathsf{Z}_2$ and the auxiliary quantum model, which describes the high-energy degrees of freedom in the mean-field approximation, is given by the transverse-field Ising model. Furthermore, in the simplest treatment, we exactly recover the result of the Gutzwiller approximation applied to the metal-insulator transition.

The outline is as follows: we first introduce the general formulation of the problem using slave-pseudospin variables in Sec.~\ref{sec:slavespinformulation}. We then discuss the mean-field approximation which consists of decoupling pseudospin and fermion degrees of freedom in Sec.~\ref{sec:SSMFT}. Furthermore, in Secs.~\ref{sec:MFA} and \ref{sec:Schwinger} we investigate the consequences of two subsequent approximations made for the pseudospin problem: (i) the single-site mean-field approximation and (ii) the use of Schwinger bosons to treat the fluctuations around the (renormalized) classical ground state. In Sec.~\ref{sec:improving} we discuss the importance to include an averaged local constraint.

\section{Slave-spin formulation}
\label{sec:slavespinformulation}
We begin this section by introducing the general framework of the slave-spin formulation we want to utilize in the study of the Hubbard model, Eq.~\eqref{eq:Hhubb}. We define a representation of physical operators in an enlarged local Hilbert space and we analyze the additional local symmetry which is introduced by this procedure. The subsequent discussion of the non-interacting case allows us to set the stage for the mean-field study in the remainder of the paper. We conclude this section by commenting on Elitzur's theorem (stating the impossibility to break the aformentioned local symmetry) and the restrictions it poses on the interpretation of the mean-field results discussed later.

For concreteness, we shall consider the single band Hubbard model written in the form
\begin{equation}
H=-\sum_{i,j,\sigma}t_{ij}c_{i\sigma}^{\dag}c_{j\sigma}+\frac{U}{2}\sum_i(\hat{n}_i-1)^2.
\label{eq:Hhubb}
\end{equation}
The hopping amplitude between sites $i$ and $j$ is denoted by $t_{ij}$ and $U$ is the onsite repulsion. $c_{i\sigma}^{(\dag)}$ destroys (creates) an electron at site $i$ with spin $\sigma$ and $\hat{n}_i=\sum_{\sigma}c_{i\sigma}^{\dag}c_{i\sigma}^{}$. Throughout the paper we work at half filling, thus assuming $\langle\hat{n}_i\rangle=1$.

\begin{figure}
\centering
\includegraphics[width=0.99\linewidth]{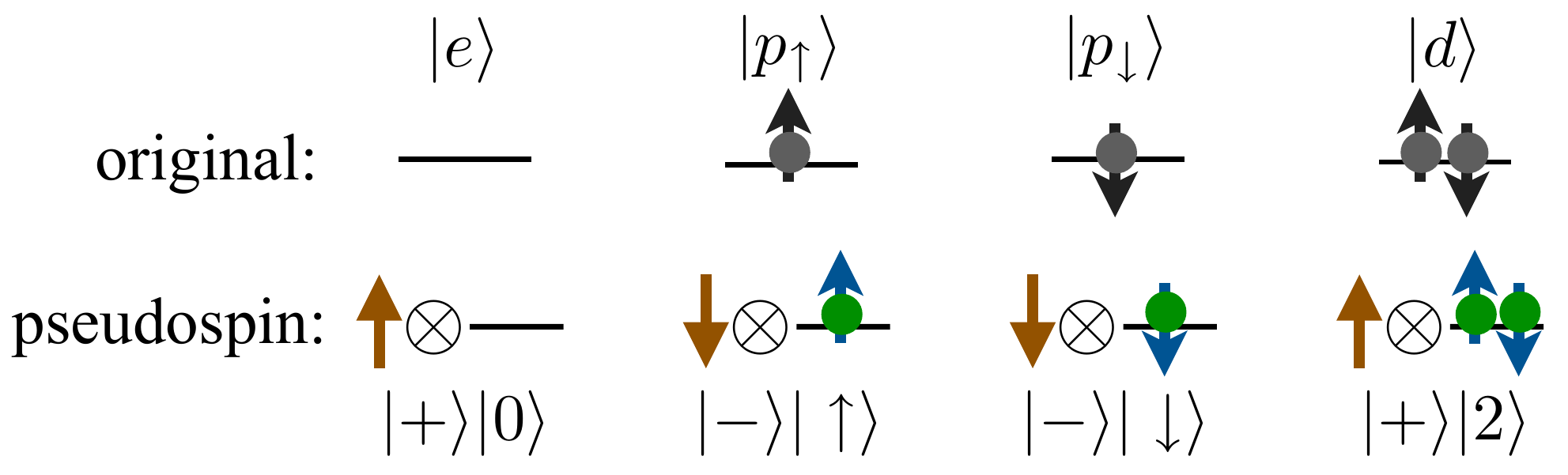}
\caption{(color online) Pictorial illustration of the pseudospin representation discussed in this article.}
\label{fig:ssrep}
\end{figure}

We now introduce a representation of the local physical states which is based on auxiliary pseudospin variables, see Fig.~\ref{fig:ssrep}. Thus, let us introduce a pseudospin ${\bf I}$ with eigenstates
\begin{equation}
I^z|\pm\rangle=\pm\frac{1}{2}|\pm\rangle,
\label{eq:pm}
\end{equation}
encoding doubly occupied and empty sites ($|+\rangle$) and singly occupied sites ($|-\rangle$). Consequently, the eigenvalue of $I^z$ refers to the presence ($-1/2$) or absence ($+1/2$) of a local magnetic moment. In addition, auxiliary Fermi creation and annihilation operators $f_{\sigma}^{(\dag)}$ are introduced to preserve the canonical anti-commutation relations (see App.~\ref{app:relation} for a connection to earlier work). The physical creation (annihilation) operator of the original model is then represented as
\begin{equation}
c_{\sigma}^{(\dag)}\equiv 2I^xf_{\sigma}^{(\dag)}.
\label{eq:If}
\end{equation}
The physical states in the enlarged local Hilbert-space are
 \begin{equation}
 |e\rangle=|+\rangle|0\rangle,\quad |p_{\sigma}\rangle=|-\rangle|\sigma\rangle,\quad |d\rangle=|+\rangle|2\rangle,
 \label{eq:physstat}
 \end{equation} 
where $|0\rangle$ is the vacuum of the $f$-fermions,
\begin{equation*}
|\sigma\rangle=f_{\sigma}^{\dag}|0\rangle\quad{\rm and}\quad|2\rangle=f_{\uparrow}^{\dag}f_{\downarrow}^{\dag}|0\rangle.
\end{equation*}
The states (\ref{eq:physstat}) are pictorially shown in Fig.~\ref{fig:ssrep}. 
In the lattice system the above definitions are generalized for each lattice site $i$.
The physical subspace is selected by the requirement that the local quasiparticle charge {\em modulo two} can be represented by $I^z_i$ as follows:
\begin{equation}
I^z_i+\frac{1}{2}-\left(n_i-1\right)^2=0.
\label{eq:conI}
\end{equation}
Associated with the constraint \eqref{eq:conI}, let us define an operator
\begin{equation*}
Q_i:=\left[I_i^z+\frac{1}{2}-\left(n_i-1\right)^2\right]^2=\frac{1}{2}+I_i^z\left[1-2(n_i-1)^2\right].
\end{equation*}
$Q_i$ has eigenvalues 0 and 1 and the eigensectors define the local physical subspace $\mathcal{H}_i^{(0)}$ and its orthogonal complement $\mathcal{H}_i^{(1)}$, respectively. Thus, the local Hilbert space $\mathcal{H}_i$ is decomposed according to 
\begin{equation*}
\mathcal{H}_i=\mathcal{H}_i^{(0)}\oplus\mathcal{H}_i^{(1)}.
\end{equation*}
We have defined
\begin{equation*}
\mathcal{H}_i^{(q)}=\left\{|\psi\rangle\in\mathcal{H}_i;\, Q_i|\psi\rangle=q|\psi\rangle\right\},\quad q=0,1.
\end{equation*}
The projection onto the physical subspace can then also be achieved by imposing for each lattice site

\begin{equation}
Q_i=0,\quad\forall\, i.
\label{eq:conQ}
\end{equation}
As a result, we can write the Hubbard interaction in the physical subspace solely using pseudospin operators $I_i^z$ and the original Hamiltonian is represented as
\begin{equation}
H'=-4\sum_{i,j,\sigma}t_{ij}I^x_iI^x_jf_{i\sigma}^{\dag}f_{j\sigma}^{}+\frac{U}{2}\sum_i\Big(I_i^z+\frac{1}{2}\Big).
\label{eq:Hprime}
\end{equation}
As long as the constraint (\ref{eq:conQ}) is fulfilled, the Hamiltonian (\ref{eq:Hprime}) is equivalent to the original model (\ref{eq:Hhubb}). In the following we often drop the constant term $U N_s/2$ ($N_s=$ number of sites) in Eq.~\eqref{eq:Hprime}. Using $Q_i$, the projector onto the physical subspace takes the form
\begin{equation}
\mathcal{P}=\prod_i(1-Q_i),
\label{eq:P}
\end{equation}
with $\mathcal{P}^2=\mathcal{P}$ and $\mathcal{P}H'\mathcal{P}$ is equivalent to the original model \eqref{eq:Hhubb}.
\subsection{Gauge structure}
In the following section, we will briefly discuss the gauge freedom introduced by the representation \eqref{eq:physstat}. Although the gauge structure is not important for the mean-field analysis outlined in the remainder of this article, it allows us to gain a deeper understanding of the slave-spin construction.

The gauge group is formed by all local transformations which leave the local physical annihilation and creation operators $c^{(\dag)}_{i\sigma}=2I_i^xf_{i\sigma}^{(\dag)}$ (and therefore also $H'$) invariant. We will see that the most obvious $\mathsf{Z}_2$-transformations
\begin{eqnarray}
c_{i\sigma}^{}\equiv 2I_i^xf_{i\sigma}^{}&=&2(-I_i^x)(-f_{i\sigma}^{}),\nonumber\\
c_{i\sigma}^{\dag}\equiv2I_i^xf_{i\sigma}^{\dag}&=&2(-I_i^x)(-f_{i\sigma}^{\dag}),
\label{eq:gauge}
\end{eqnarray}
are in fact {\it not} the most general ones but are part of a larger $\mathsf{U}(1)$ group. Nevertheless, as shown below, the $\mathsf{Z}_2$-gauge transformations \eqref{eq:gauge} are the only ones respecting the mean-field decomposition and in this sense, the mean-field ansatz breaks the $\mathsf{U}(1)$ gauge symmetry down to $\mathsf{Z}_2$.

In order to explicitly derive all possible gauge transformations we make use of the fact that $H'$ does not mix the physical subspace with its complement. The generalized local charge $Q_i$ is therefore a conserved quantity and commutes with $H'$:
\begin{equation*}
[Q_i,H']=0.
\end{equation*}
A conserved charge $Q_i$ arises from a continuous symmetry $U_i(\phi_i)=\exp(i\phi_i\, Q_i)$ with $0\leq\phi_i<2\pi$. In other words, $Q_i$ generates the local gauge transformations $U_i(\phi_i)$ and we conclude that there is a local $\mathsf{U}(1)$ freedom,
\begin{equation}
U_i(\phi_i)c_{i\sigma}^{(\dag)}U_i(\phi_i)^{\dag}=c^{(\dag)}_{i\sigma},
\label{eq:ctilde}
\end{equation}
with $c^{(\dag)}_{i\sigma}=2I_i^xf_{i\sigma}^{(\dag)}$. Under the action of $U_i(\phi_i)$ the pseudo-fermions transform as
\begin{equation}
U_i(\phi_i)f_{i\sigma}^{(\dag)}U_i(\phi_i)^{\dag}=e^{i\phi_i}f_{i\sigma}^{(\dag)}(1-Q_i)+e^{-i\phi_i}f_{i\sigma}^{(\dag)}Q_i,
\label{eq:ftilde}
\end{equation}
and the pseudospins as
\begin{eqnarray}
U_i(\phi_i)I_i^xU_i(\phi_i)^{\dag}&=&e^{i\phi_i}I_i^x(1-Q_i)+e^{-i\phi_i}I_i^xQ_i,\nonumber\\
U_i(\phi_i)I_i^zU_i(\phi_i)^{\dag}&=&I_i^z.
\label{eq:Itilde}
\end{eqnarray}
The transformation relations \eqref{eq:ftilde} and \eqref{eq:Itilde} allow to explicitly check the relation \eqref{eq:ctilde}.

At this point it is important to recall that the present slave-spin scheme is a projective construction: a physical state $|\Psi\rangle$ is obtained from a general state $|\Phi\rangle$ of the enlarged Hilbert space after projection, $|\Psi\rangle=\mathcal{P}|\Phi\rangle$. A different state $|\Phi'\rangle$ which is obtained from $|\Phi\rangle$ by a gauge transformation will give rise to the {\it same} physical state $|\Psi\rangle$. (For a related statement in the projective construction of spin liquid phases see e.g. Refs.~\onlinecite{Wen:1991} and \onlinecite{Wen:2002}.) This is an intuitive way to understand the origin of the gauge freedom. In practice, we start from a mean-field (product) state in pseudospin and pseudo-fermion degrees of freedom: $|\Phi_{\rm MF}\rangle=|\Phi_I\rangle|\Phi_f\rangle$, see Sec.~\ref{sec:SSMFT}. Applying the transformations \eqref{eq:ftilde} and \eqref{eq:Itilde} to the mean-field Hamiltonians \eqref{eq:Hf} and \eqref{eq:HI} we find that in general only gauge transformations with $\phi_i=0$ or $\pi$ respect the mean-field product form. This means that different {\it mean-field states} which give rise to the same physical state are related by gauge transformations with $\phi_i=0$ or $\pi$. In this sense, the mean-field ansatz breaks the $\mathsf{U}(1)$ freedom down to the smaller $\mathsf{Z}_2$ freedom. Therefore, we anticipate that the relevant gauge group to study fluctuations around the mean-field state is $\mathsf{Z}_2$. It consists of the following two elements: the identity operator, $\mathsf{id}_i\equiv U_i(\phi_i=0)$, and the operator 
\begin{equation}
u_i\equiv U_i(\phi_i=\pi)=(-1)^{Q_i}=1-2 Q_i.
\label{eq:ui}
\end{equation}
The action of $u_i$ on $f_{i\sigma}^{(\dag)}$ and $I_i^x$ is given by
\begin{eqnarray*}
u_if_{i\sigma}^{(\dag)}u_i&=&-f_{i\sigma}^{(\dag)},\\
u_iI_i^x u_i&=&-I_i^x.
\end{eqnarray*}
Obviously, $u_i$ promotes the gauge transformations \eqref{eq:gauge}. 

\subsection{Non-interacting model $U=0$}
We now turn to the discussion of the non-interacting model:
\begin{equation*}
H_0'=-4\sum_{i,j,\sigma}t_{ij}I^x_iI^x_jf_{i\sigma}^{\dag}f_{j\sigma}^{}.
\end{equation*}
The eigenstates of $H_0'$ are (at least in principle) exactly known. This fact allows us to study the action of the projector $\mathcal{P}$ and the role of the local gauge freedom in a more rigorous manner. Moreover, the non-interacting limit is helpful for the interpretation of the mean-field approximation introduced and explored later in this article. In the following we use arguments which are similar to those given in a discussion of an exactly solvable spin-model on the square lattice in Ref.~\onlinecite{Yao:2009}. 

Since $[I_i^x,H_0']=0$ we choose the eigenstates $|\Phi\rangle$ of $H_0'$ (in the enlarged Hilbert space) as being product states in the (Ising-) pseudospin and fermion degrees of freedom. Explicitly, $|\Phi\rangle=|\{\alpha\}\rangle|\phi(\{\alpha\})\rangle$ where $I_i^x|\{\alpha\}\rangle=\alpha_i|\{\alpha\}\rangle$, $\alpha_i=\pm1/2$. The pseudo fermion component $|\phi(\{\alpha\})\rangle$ is a Slater-Determinant obtained from the non-interacting model with effective hopping parameters $4\alpha_it_{ij}\alpha_j$. As an example, let us choose $\alpha_i=+1/2$ for all sites $i$. The ground state in this sector is given by $|\Phi_0\rangle=|\{1/2\}\rangle|\phi_0(\{1/2\})\rangle$ where $|\phi_0(\{1/2\})\rangle$ is the Fermi sea of the pseudo fermions. Clearly, $|\Phi_0\rangle$ has the same energy $E_0$ as the physical ground state:
\begin{equation*}
H_0'|\Phi_0\rangle=E_0|\Phi_0\rangle.
\end{equation*}
However, the state $|\Phi_0\rangle$ is not an eigenstate of the projector $\mathcal{P}$ and, therefore, is not the physical ground state. Instead, the (unnormalized) physical ground state is obtained by projecting $|\Phi_0\rangle$ onto the physical subspace, $|\Psi_0\rangle=\mathcal{P}|\Phi_0\rangle$.\cite{note:P}

At this point it is important to note that the local $\mathsf{Z}_2$ gauge freedom \eqref{eq:gauge}, $[H_0',u_i]=0$, implies that each energy sector of $H_0'$ is macroscopically degenerate $(\sim 2^{N_s})$. In fact, applying any combination of local gauge transformations $u_i$ to $|\Phi_0\rangle$ yields an eigenstate of $H_0'$ with energy $E_0$ but which is (in general) orthogonal to $|\Phi_0\rangle$. This property allows to obtain the physical ground state by a particular superposition of eigenstates in the $E_0$-sector. To see this, let us take a closer look at the projector. Using Eq.~\eqref{eq:ui} the projector Eq.~\eqref{eq:P} is written in the form
\begin{equation}
\mathcal{P}=\prod_i\frac{1+u_i}{2}.
\end{equation}
Explicitly, it takes the form
\begin{equation}
\mathcal{P}=\left(1+\sum_iu_i+\sum_{i_1<i_2}u_{i_1}u_{i_2}+\dots+\prod_iu_i\right)/2^{N_s}.
\label{eq:Pui}
\end{equation}
The action of $u_i$ on an eigenstate $|\Phi\rangle$ of $H_0'$ can be understood by writing $u_i$ in the form
\begin{equation}
u_i=-2I_i^z[1-2(n_i-1)^2].
\label{eq:ui-exp}
\end{equation}
Therefore, $u_i$ changes the sign of $\alpha_i$ in the pseudospin component $|\{\alpha\}\rangle$. To understand the action of $u_i$ on the pseudo fermion part, we expand $|\phi(\{\alpha\})\rangle$ in the site diagonal occupation number basis. The form \eqref{eq:ui-exp} of $u_i$ implies that the components with an empty or doubly occupied site $i$ are multiplied by $-1$ while those with a singly occupied site $i$ are not changed. The resulting wave-function is then just the corresponding eigenstate of a non-interacting model in which $f_{i\sigma}^{(\dag)}$ is replaced by $-f_{i\sigma}^{(\dag)}$. Thus, $u_i$ indeed acts as a local gauge transformation:
\begin{equation}
u_i\left(|\{\alpha\}\rangle|\phi(\{\alpha\})\rangle\right)=-|\{\alpha'\}\rangle|\phi(\{\alpha'\})\rangle,
\end{equation}
where $\{\alpha'\}=\{\dots,\alpha_{i-1},-\alpha_i,\alpha_{i+1},\dots\}$. Hence, owing to the form \eqref{eq:Pui} of $\mathcal{P}$, the physical ground state $|\Psi_0\rangle=\mathcal{P}|\Phi_0\rangle$ is the equal amplitude superposition of all the degenerate states which are obtained from $|\Phi_0\rangle$ by applying all possible local gauge transformations.\cite{Yao:2009} The physical subspace is therefore the gauge invariant subspace.

Note that, although $|\Phi_0\rangle$ is not the physical ground state, it is a {\it representative state of the ground-state energy sector} and, as long as gauge invariant operators $\hat{O}=u_i\hat{O}u_i$ are considered, the expectation values are equal, $\langle\Phi_0|\hat{O}|\Phi_0\rangle=\langle\Psi_0|\hat{O}|\Psi_0\rangle$.\cite{Kogut:1979} Because physical observables are gauge invariant we can calculate all physical properties of the non-interacting model by restricting to the sector $\alpha_i=1/2$ for all $i$ (or to any other fixed configuration $\{\alpha\}$).
\subsection{Relevance of Elitzur's theorem}
\label{subsec:Elitzur}
In this paragraph, we briefly comment on Elitzur's theorem\cite{Elitzur} which states the impossibility to spontaneously break a local symmetry. This implies that thermal averages $\langle\dots\rangle_{\rm th}$ in the physical as well as in the enlarged Hilbert space of operators which are not gauge invariant have to be zero. In the present case, it implies, for example, that $\langle I_i^x\rangle_{\rm th}=0$.\cite{note:proof}

In the non-interacting limit discussed above this result is plausible. Let us first consider thermal averages in the enlarged Hilbert space: because each energy sector of $H'_0$ is spanned by a macroscopic number of states involving all possible pseudospin configurations $\{\alpha\}$ the expectation value of $I_i^x$ averages to zero. Likewise, restricting the thermal average to the physical subspace, the fact that physical states are equal amplitude superpositions of all the degenerate states in the enlarged Hilbert space results in a mutual cancellation of positive and negative contributions to $\langle I_i^x\rangle_{\rm th}$. In particular, for the physical ground state we find $\langle\Psi_0|I_i^x|\Psi_0\rangle=0$.

However, the expectation value of a single state in the enlarged Hilbert space can have a non-vanishing expectation value $\langle\Psi|I_i^x|\Psi\rangle\neq0$. Consider for example the exact eigenstate $|\Phi_0\rangle$ of $H_0'$ defined in the previous section. Although $\langle\Phi_0|I_i^x|\Phi_0\rangle=1/2$ we can correctly obtain all the physical observables in the ground state from $|\Phi_0\rangle$ because it is a representative state of the physical ground-state sector. 

We would like to argue that the mean-field approximation introduced in the next section should be interpreted in the same sense. Namely, in the mean-field approximation, we seek for a product state which approximates one particular ground state of $H'$ in the enlarged Hilbert space. In fact, for small $U$, the mean-field state is continuously connected to $|\Phi_0\rangle$. While this procedure does not yield a systematic approximation we assume that also for larger $U$ the mean-field state is sufficiently close to a representative state of the ground-state energy sector. Moreover, the mean-field approximation is truly variational: it gives an upper limit for the true ground state energy of the original model. Although a mean-field state will in general have $\langle I_i^x\rangle\neq0$, we can use it to approximatively calculate physical ground state properties. 

We think that a similar interpretation of other slave-particle mean-field theories is appropriate. This would be consistent with observations made by studying slave-boson theories in the ``radial gauge" where all the exact physical properties of a toy model have been obtained from the saddle-point solution of a functional integral.\cite{Fresard:2007,Fresard:2008}
\section{Mean-field theory}
\label{sec:SSMFT}
Let us now discuss the mean-field theory which results from the representation introduced in the previous section. To this end, we assume product states $|\Psi\rangle=|\Psi_I\rangle|\Psi_f\rangle$ in pseudospin and fermion degrees of freedom. These states live in the enlarged Hilbert space. As discussed in the previous section, we assume that $|\Psi\rangle$ yields a sufficiently good approximation to a state in the (macroscopically degenerate) ground-state sector of $H'$ in the enlarged Hilbert space. We postpone a discussion of the importance to include the constraint \eqref{eq:conI} on average to Sec.~\ref{sec:improving}. Note, however, that the relation
\begin{equation}
\langle I_i^z\rangle+\frac{1}{2}=\frac{1}{N_s}\frac{\partial}{\partial U}\langle H'\rangle\equiv \Big\langle \Big(\sum_{\sigma} c_{i\sigma}^{\dag} c_{i\sigma}^{}-1\Big)^2\Big\rangle,
\label{eq:IvsD}
\end{equation}
holds, which leads to the identification of the physical fraction of doubly occupied sites, $\langle c_{i\uparrow}^{\dag}c_{i\uparrow}^{}c_{i\downarrow}^{\dag}c_{i\downarrow}^{}\rangle$, with $\langle I_i^z\rangle$/2+1/4. This is analogous to earlier mean-field treatments, see Eq.~\eqref{eq:conD}.
The rational behind the mean-field decoupling is the fact that we can approximately distinguish two energy scales. Indeed, as shown below, the characteristic scale of the pseudospins is  $\sim\max (U,U_c)$ whereas that of the pseudo-fermions is $\sim t$ [$U_c$ is given in Eq.~\eqref{eq:Jbar}]. Although this observation together with the insights gained from the non-interacting limit justifies to some extent the mean-field decoupling, it should be considered as a first step on which a more sophisticated analysis can be based. Nevertheless, on the present level of approximations we can make close contact to earlier results.
\subsection{Mean-field Hamiltonians}
As a consequence of the mean-field decoupling we obtain two effective Hamiltonians: the fermion problem assumes the form of a non-interacting tight-binding Hamiltonian,
\begin{equation}
H_f=\langle\Psi_I|H'|\Psi_I\rangle=-\sum_{i,j,\sigma}g_{ij}t_{ij}f_{i\sigma}^{\dag}f_{j\sigma}^{},
\label{eq:Hf}
\end{equation}
with the hopping amplitude $t_{ij}$ renormalized by a factor 
\begin{equation}
g_{ij}=4\langle I_i^xI_j^x\rangle_I.
\label{eq:gij}
\end{equation} 
On the other hand, the pseudospin problem reduces to the quantum Ising model
\begin{equation}
H_I=\langle\Psi_f|H'|\Psi_f\rangle=-\sum_{(i,j)}\chi_{ij}t_{ij}I_i^xI_j^x+\frac{U}{2}\sum_iI_i^z,
\label{eq:HI}
\end{equation}
with the transverse field $U/2$ and the Ising exchange coupling $t_{ij}\chi_{ij}$ where
\begin{equation}
\chi_{ij}=4\sum_{\sigma}\left(\langle f_{i\sigma}^{\dag}f_{j\sigma}\rangle_f+{\rm c.c}\right).
\label{eq:chiij}
\end{equation}
The sum over $(i,j)$ in Eq.~(\ref{eq:HI}) means summation over all bonds $(i,j)$ where $\chi_{ij}\neq 0$.

The quantum Ising model (\ref{eq:HI}) is a prime example of a system displaying a quantum critical point at a critical ratio of the transverse field to the Ising coupling. It  separates a magnetically ordered from a quantum paramagnet.\cite{sachdev:1999} Below we show that this quantum critical point can be identified with the Brinkman-Rice criticality of an almost localized Fermi liquid.\cite{Brinkman:1970,Vollhardt:1984} 

Equations (\ref{eq:gij}) and (\ref{eq:chiij}) are the two coupled self-consistency equations to be solved in the mean-field approximation. We note that there is always the trivial solution $g_{ij}=\chi_{ij}=0$ of these equations which is the physical solution in the atomic limit $t_{ij}=0$. Non-trivial solutions invoke an (approximate) solution of the quantum Ising model.

The symmetry properties of the original model are conserved in the mean-field approximations \eqref{eq:Hf} und \eqref{eq:HI}. For example, particle number conservation leads to a global $\mathsf{U}(1)$ symmetry for the slave fermion sector [Eq.~\eqref{eq:Hf}] in the usual manner. Within our approach, the pseudospin sector explicitly breaks the pseudospin-rotation symmetry [Eq.~\eqref{eq:HI}]. This is {\em not} related to the particle number, however, because the pseudospins measure charge only {\em modulo 2}. Therefore, the appearance of terms $\propto I_i^xI_j^x$ (instead of $I_i^+I_j^-$) do not bias our system toward a $\mathsf{U}(1)$ symmetry broken phase in the mean-field approximation.
\subsection{Single-particle Green function}
Eventually, we are interested in physical (gauge-invariant) quantities such as the single-particle Green function,
\begin{equation*}
G_{\sigma}({\bf r}_i,{\bf r}_j;t)=-i\langle T c_{j\sigma}(t)c_{i\sigma}^{\dag}(0)\rangle,
\end{equation*}
where $T$ denotes time ordering. In the mean-field theory, $G_{\sigma}$ is obtained as
\begin{equation}
G_{\sigma}({\bf r}_i,{\bf r}_j;t) \approx 4B_{ij}(t)G_{\sigma}^{f}({\bf r}_i,{\bf r}_j;t),
\label{eq:Gmf}
\end{equation}
where we have introduced the auxiliary quantities
\begin{eqnarray*}
B_{ij}(t)&=&\langle TI_{j}^x(t)I_{i}^{x}(0)\rangle,\\
G_{\sigma}^{f}({\bf r}_i,{\bf r}_j;t)&=&-i\langle Tf_{j\sigma}(t)f_{i\sigma}^{\dag}(0)\rangle.
\end{eqnarray*}
In momentum and energy space, the relation \eqref{eq:Gmf} translates into a convolution of $B({\bf q},\omega)$ and $G_{\sigma}^f({\bf q},\omega)$. It is noteworthy to mention that the canonical anti-commutation relations of the physical annihilation and creation operators are preserved on average,
\begin{equation}
\langle\{c_{i\sigma}^{},c_{j\sigma'}^{\dag}\}\rangle=4\langle I_i^xI_j^x\rangle\langle\{f_{i\sigma}^{},f_{j\sigma'}^{\dag}\}\rangle=\delta_{ij}\delta_{\sigma\sigma'},
\label{eq:anti}
\end{equation}
where $\langle\dots\rangle$ denotes the average over mean-field eigenstates. As a consequence, the single-particle spectral density
is correctly normalized, as long as the spin-1/2 identity $\left(I_i^x\right)^2=1/4$ is respected. 

\section{Mean-field approximations to the Ising model}
\label{sec:MFA}
A straight forward way to study the quantum Ising model (\ref{eq:HI}) is mean-field approximations. In the simplest case we consider a single pseudospin coupled to a self-consistent effective field. This is the local approximation discussed in Sec.~\ref{subsec:local}. In order to improve over the local approximation we consider in Secs.~\ref{subsec:cluster} and \ref{subsec:FS} the pseudospin problem on a finite cluster; see Refs.~\onlinecite{Zhao:2007} and \onlinecite{Hassan:2010} for related works. This allows us to discuss important aspects of inter-site correlations which are absent in the local approximation. In the following we work in the zero temperature limit and we restrict our analysis to translation-invariant and paramagnetic states. 

\subsection{Local approximation}
\label{subsec:local}
We start with the single-site ``cluster". Noteworthy, on this level of the approximation, the mean-field self-consistency (\ref{eq:gij}) and (\ref{eq:chiij}) leads to the Brinkman-Rice transition\cite{Brinkman:1970} obtained in the paramagnetic Gutzwiller approximation of the Hubbard model.\cite{Gutzwiller:1963} To see this, let us introduce the mean magnetization $\langle I_0^x\rangle$ and
\begin{equation}
H_I^{MF}=h\sum_i\tilde{I}^z_i,\quad h=\frac{U_c}{2}\sqrt{u^2+4\langle I_0^x\rangle^2},
\label{eq:HIMF}
\end{equation}
where the pseudospin has been rotated due to the action of the mean field $U_c\langle I_0^x\rangle$,
\begin{equation}
\tilde{{\bf I}}_i=e^{i\alpha I^y_i}{\bf I}e^{-i\alpha I^y_i},\quad\tan\alpha=\frac{2\langle I_0^x\rangle}{u}.
\label{eq:rotI}
\end{equation}
In Eqs.~(\ref{eq:HIMF}) and (\ref{eq:rotI}) we have used the dimensionless interaction parameter $u=U/U_c$ where the energy scale $U_c$ is associated with the pseudospin Ising coupling in the single-site solution
\begin{equation}
U_c=\sum_jt_{ij}\chi_{ij}=-16 \int_{-D}^{\varepsilon_F}d\varepsilon\varepsilon\rho_{\sigma}(\varepsilon)>0.
\label{eq:Jbar}
\end{equation}
Here, $\rho_{\sigma}(\varepsilon)$ is the non-interacting density of states per spin with half bandwidth $D$ and $\varepsilon_F$ is the effective Fermi energy of the pseudo fermions. 

Self-consistency of $\langle I_0^x\rangle$ yields the pseudospin magnetization and from Eq.~(\ref{eq:gij}) we obtain the hopping renormalization factor as follows:
\begin{equation}
g=4\langle I_0^x\rangle^2=
\begin{cases}
1-u^2, & u\leq 1;\\
0,&u>1.
\end{cases}
\label{eq:GA}
\end{equation}
In particular, on this level of the approximation, the effective mass $m^*/m=1/g$ diverges at the critical interaction strength $U_c$, indicating the transition to the localized state. The double occupancy follows from the relation (\ref{eq:IvsD}):
\begin{equation}
d^2=\frac{1}{2}\left(\langle I_0^z\rangle+\frac{1}{2}\right)=\begin{cases}
\frac{1-u}{4}, & u\leq1;\\
0, & u>1.
\end{cases}
\label{eq:GAd}
\end{equation}
The angle $\alpha$ of the pseudospin rotation (\ref{eq:rotI}) can be written in terms of $d$ as
\begin{equation}
\cos\frac{\alpha}{2}=\sqrt{1-2d^2}.
\label{eq:alphad}
\end{equation}
Following Eq.~(\ref{eq:alphad}), the rotation of the pseudospin corresponds to adjusting the average fraction $d^2$ of doubly-occupied sites. 

Note that the single-site solution (\ref{eq:GA}) and (\ref{eq:GAd}) for $u>1$ reproduces the result of the atomic limit $t_{ij}=0$. Due to the fact that inter-site correlations have been neglected the low-energy physics of the Mott insulator is completely absent. This shortcoming can be addressed by going beyond the local approximation.

Our mean-field state has a finite expectation value $\langle I_i^x\rangle\neq 0$ for $u<1$. However, according to Elitzure's theorem, the physical ground state requires $\langle I_i^x\rangle=0$ because $I_i^x$ is not a gauge invariant operator. As we have argued in Sec.~\ref{subsec:Elitzur} we consider our mean-field state as an approximation to a representative eigenstate in the enlarged Hilbert space. In this light, we can still expect that reasonable approximations for physical (gauge invariant) observables are obtained.
\subsection{Cluster approximations}
\label{subsec:cluster}
To improve over the local approximation we have studied clusters with two or more sites. We briefly review the generic differences. A particular example is discussed in the next subsection. Most important is the fact that in finite spatial dimensions the value of $\langle I_i^x\rangle^2$ and  
$\langle I_i^xI_j^x\rangle$ with $i\neq j$ are different. In other words, inter-site correlations are introduced. Consequently, in accordance with previous studies,\cite{Florens:2004, Hassan:2010} there is a distinction between the quasiparticle weight $Z$ and the effective mass renormalization $m/m^*$ of the quasiparticles. While in general the factor $g=m/m^*$ stays finite across the Mott transition, the quasiparticle weight $Z$ still vanishes for $U\rightarrow U_c$. The quasiparticle weight and the effective mass in the mean-field theory follow from the form \eqref{eq:Gmf} of the single-particle Green's function. For a nearest-neighbor hopping model they are given by 
\begin{equation*}
Z=4\langle I_i^x\rangle^2\quad\mathrm{and}\quad\frac{m}{m^*}=g=4\langle I_i^xI_j^x\rangle,
\end{equation*}
for nearest-neighbor pairs $i,j$. The distinction between $g$ and $Z$ is now apparent in finite dimensions. More general, the electronic self-energy obtains a ${\bf k}$ dependence which, in the metallic phase at particle-hole symmetry, is of the form
\begin{equation*}
\Sigma(\omega,{\bf k})=(1+Z^{-1})\omega+(\frac{g}{Z}-1)\varepsilon_{\bf k}+\dots
\end{equation*} 
for ${\bf k}$ near the Fermi surface defined by $\varepsilon_{\bf k}=0$ and for small $\omega$'s. This issue will again be discussed in Sec.~\ref{sec:Schwinger} when we consider the Schwinger boson mean-field theory in order to access the low lying excitations of the quantum Ising model. 

In addition, the critical interaction strength for the Mott transition is renormalized compared to the value of the local approximation (\ref{eq:Jbar}). We find, however, that the exact value depends on the choice of the cluster.
\subsection{Deformation of Fermi surface}
\label{subsec:FS}
\begin{figure}
\centering
\includegraphics[width=0.9\linewidth]{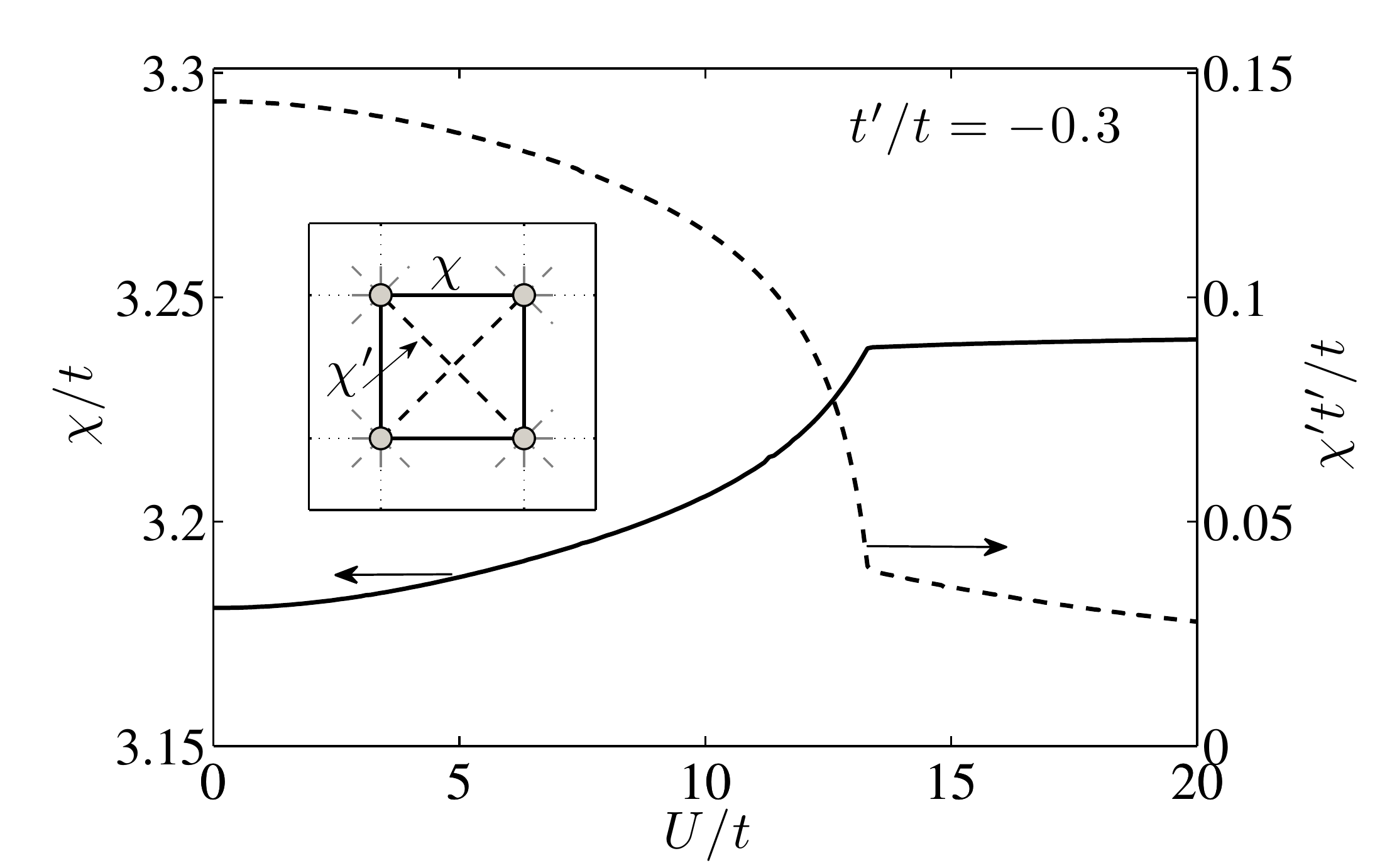}
\caption{The nearest-neighbor exchange $\chi t$ and the next-nearest neighbor exchange $\chi' t'$ of the auxiliary pseudospin model as function of $U/t$ for $t'=-0.3t$. The inset shows the considered cluster.}
\label{fig:JJp}
\end{figure}

\begin{figure}
\centering
\includegraphics[width=0.9\linewidth]{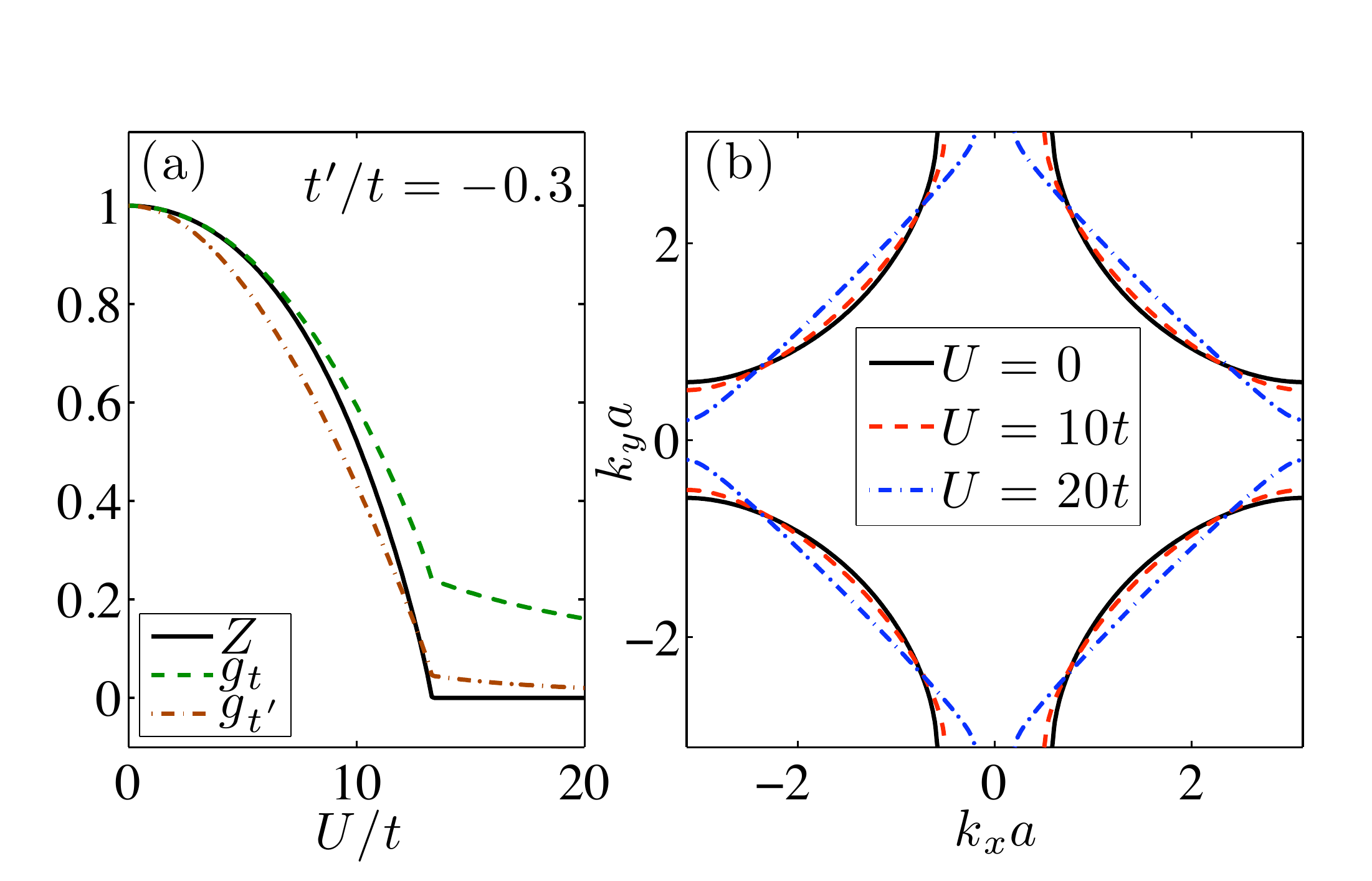}
\caption{(color online) (a) The quasiparticle weight $Z$ and the nearest-neighbor and next-nearest-neighbor hopping renormalization factors $g_t$ and $g_{t'}$, respectively, as function of $U/t$. (b) The Fermi surface for different values of the interaction strength.}
\label{fig:gsFS}
\end{figure}

As an application of the cluster mean-field scheme, let us now study the question how inter-site correlations can change the shape of the Fermi surface. Such an interaction-driven deformation is easily described within the present scheme when considering a generic dispersion with further-neighbor hopping amplitudes. The deformation then results from a different renormalization of nonequivalent hopping amplitudes. As an example, we study here a model on a two-dimensional square lattice with nearest-neighbor hopping amplitude $t$ and next-nearest-neighbor hopping amplitude $t'$. The renormalized quasiparticle dispersion is then given by
\begin{equation*}
\xi_{\bf k}=-2g_t t\left(\cos k_x+\cos k_y\right)-4g_{t'}t'\cos k_x\cos k_y-\epsilon_F.
\end{equation*}
Here, the renormalization factors are $g_{t^{(')}}=4\langle I_i^xI_j^x\rangle$ with $i$ and $j$ (next-)nearest neighbors. The parameter $\epsilon_F$ is determined to satisfy the Landau-Luttinger sum rule for the Fermi surface defined by $\xi_{\bf k}=0$. According to Eq.~(\ref{eq:chiij}), the auxiliary quantum Ising model acquires the nearest neighbor exchange
\begin{equation*}
\chi t=4t\sum_{{\bf k},\sigma}(\cos k_x+\cos k_z)n_{{\bf k}\sigma},
\end{equation*}
and a next-nearest neighbor coupling
\begin{equation*}
\chi' t'=8t'\sum_{{\bf k},\sigma}\cos k_x\cos k_y n_{{\bf k}\sigma}.
\end{equation*}
We have considered a $2\times 2$ cluster as shown in the inset of Fig.~\ref{fig:JJp} and have solved the self-consistency Eqs.~(\ref{eq:gij}) and (\ref{eq:chiij}) for different values of $U/t$ with a fixed ratio $t'/t=-0.3$. Figure~\ref{fig:JJp} shows $\chi$ and $\chi'$ as function of $U/t$. Note the difference in scale for $\chi$ and $\chi'$.
Figure~\ref{fig:gsFS}(a) shows the quasiparticle weight $Z$ along with $g_t$ and $g_{t'}$. As mentioned in the previous subsection, $Z$ vanishes in the Mott insulator while $g_{t}$ and $g_{t'}$ stay finite. On the metallic side, $g_{t'}<g_t$ which means that the Fermi surface is deformed toward the fully nested surface with $t'=0$, as shown in Fig.~\ref{fig:gsFS}(b). A similar behavior has been found in calculations which take antiferromagnetic fluctuations into account.\cite{Yanase:1999} On the insulating side, $g_{t'}\ll g_t$. This can be expected since $g_{t'}/g_t\sim J'/J\sim (t'/t)^2$ in the Mott insulator where $J$ $(J')$ is the super-exchange for (next-) nearest neighbors. For $U>U_c$, the low-lying excitations are not given by Landau quasiparticles since $Z=0$. The surface for $U=20t$ shown in panel (b) can be viewed as the Fermi surface of a ``hidden Fermi liquid" \cite{Anderson:2008} or as the spinon Fermi surface of a gapless spin liquid.\cite{Lee:2006b} However, since this Fermi surface is close to perfect nesting we expect that (residual) interactions open a gap. Before we study the effect of such residual interactions in Sev.~\ref{sec:improving} in more detail, we present another way of going beyond the local approximation in the next section.
%
\section{Schwinger boson theory}
\label{sec:Schwinger}
An alternative approach to go beyond the local approximation is the use of the Schwinger boson theory for the quantum Ising model and we follow an essentially similar line of thoughts as in Ref.~\onlinecite{Huber:2007}. The bosonic representation establishes also a connection to earlier slave-boson representations, see App.~\ref{app:relation}. Moreover, it allows us to take the effect of quantum fluctuations into account and it yields the low-lying excitations. 

We introduce two sets of Bose creation and annihilation operators $x_i^{(\dag)}$ and $y_i^{(\dag)}$ to represent the pseudospin algebra
\begin{equation*}
I^+_i=y_i^{\dag}x_i,\quad I_i^-=x_i^{\dag}y_i,\quad I^z_i=y_i^{\dag}y_i-\frac{1}{2}.
\label{eq:SWxy}
\end{equation*}
In order to obtain a faithful representation of the spin algebra, these operators have to obey
\begin{equation*}
x^{\dag}_ix_i+y_i^{\dag}y_i=1.
\end{equation*}
In the Schwinger boson formulation, the pseudospin rotation given in Eq.~\eqref{eq:rotI} translates to an unitary transformation of the Bose creation and annihilation operators
\begin{equation}
\left(\begin{array}{c}
x_i\\ y_i
\end{array}\right)=\left(
\begin{array}{cc}
\cos\frac{\alpha}{2} & -\sin\frac{\alpha}{2}\\
\sin\frac{\alpha}{2} & \cos\frac{\alpha}{2}
\end{array}\right)\left(
\begin{array}{c}
a_i\\ b_i
\end{array}\right).
\label{eq:rotxy}
\end{equation}
The $a$ and $b$ bosons can be interpreted as the Schwinger bosons of the rotated pseudospin $\tilde{{\bf I}}_i$. Equally, using the relation (\ref{eq:alphad}), we can specify the transformation (\ref{eq:rotxy}) by the value of $d$ and in the following we denote the canonically transformed Ising model by $H_B(d)$.
\subsection{Classical ground state}
The result of the local approximation is re-obtained by assuming a product form of the wave function
\begin{equation}
|\mathcal{O}\rangle=\prod_ia_i^{\dag}|0\rangle,
\label{eq:B}
\end{equation}
and optimizing the energy 
\begin{equation*}
E(d)=\langle \mathcal{O}|H_B(d)|\mathcal{O}\rangle,
\end{equation*}
with respect to the parameter $d$. We recover again the Gutzwiller result (\ref{eq:GA}).
The state (\ref{eq:B}) is the classical ground state of the transverse Ising model. We state here that this is the point where in the KR functional integral representation one stops if only the saddle point (without fluctuations around it) is considered. 
\subsection{Role of fluctuations in three dimensions}
\label{sec:3DMott}
The formalism developed so far offers a framework to study the excitation spectrum of the transverse Ising model. We show that these excitations lead to the incoherent one-particle excitations of the Hubbard model (upper and lower Hubbard bands). Furthermore, quantum fluctuations renormalize the classical ground state obtained in the mean-field approximation. In general, the role of fluctuations crucially depends on the dimensionality of the system. Here we restrict our analysis to the three dimensional (cubic) lattice where fluctuations around the classical ground state are small for most parameters. Nevertheless, the finite dimensionality reintroduced in our analysis causes interesting changes in the nature of the Mott transition as compared to the infinite dimensional result. Our results are in agreement with previous results based on the slave-rotor formalism.\cite{Florens:2004}
\subsubsection{Effective Hamiltonian for fluctuations}
Formally, the effective Hamiltonian is derived by expanding $H_B(d)$ up to second order in the $b$ bosons. The parameter $d$ of the unitary transformation is then determined by the requirement that quadratic mixing terms in $a$ and $b$ of $H_B(d)$ vanish. This yields again the condition (\ref{eq:GAd}). The next step is to let the $a$ bosons condense and to replace the $a$ operators by unity. In momentum space the effective Hamiltonian then reads
\begin{equation}
H_{\rm eff}=\frac{U_c}{4}\sum_{\bf k}\mathcal{B}_{\bf k}^{\dag}\left(\begin{array}{cc}\frac{u^2}{2}\gamma_{\bf k}+1&\frac{u^2}{2}\gamma_{\bf k}\\ \frac{u^2}{2}\gamma_{\bf k}&\frac{u^2}{2}\gamma_{\bf k}+1\end{array}\right)\mathcal{B}_{\bf k},
\label{eq:Heff<}
\end{equation}
for $u\leq 1$. For $u>1$ we find
\begin{equation}
H_{\rm eff}=\frac{U_c}{4}\sum_{\bf k}\mathcal{B}_{\bf k}^{\dag}\left(\begin{array}{cc}\frac{1}{2}\gamma_{\bf k}+u&\frac{1}{2}\gamma_{\bf k}\\ \frac{1}{2}\gamma_{\bf k}&\frac{1}{2}\gamma_{\bf k}+u\end{array}\right)\mathcal{B}_{\bf k}.
\label{eq:Heff>}
\end{equation}
We have introduced the operators $\mathcal{B}_{\rm k}^{(\dag)}$ given by
\begin{equation*}
\mathcal{B}_{\bf k}^{\dag}=\left(b_{\bf k}^{\dag}, b_{-\bf k}\right)\quad{\rm and}\quad\mathcal{B}_{\bf k}=
\left(\begin{array}{c}
b_{\bf k}\\
b_{-\bf k}^{\dag}
\end{array}\right),
\end{equation*}
as well as
\begin{equation*}
\gamma_{\bf k}=-\frac{1}{3}\sum_{i=1}^{3}\cos(k_i).
\end{equation*}
The effective Hamiltonian for fluctuations, Eq.~(\ref{eq:Heff<}) and (\ref{eq:Heff>}), is diagonalized by the following Bogoliubov transformation
\begin{equation*}
\left(\begin{array}{c}b_{\bf k}\\ b_{-{\bf k}}^{\dag}\end{array}\right)=\left(\begin{array}{cc}\cosh\vartheta_{\bf k}&\sinh\vartheta_{\bf k}\\ \sinh\vartheta_{\bf k} & \cosh\vartheta_{\bf k} \end{array}\right)\left(\begin{array}{c}\beta_{\bf k}\\\beta_{-{\bf k}}^{\dag}\end{array}\right).
\end{equation*}
The mixing angle is given by
\begin{equation*}
\vartheta_{\bf k}=\frac{1}{2}{\rm atanh}\left[\frac{-\min(1,u^2)\gamma_{\bf k}}{\min(1,u^2)\gamma_{\bf k}+2\max(1,u)}\right].
\end{equation*}
The $\beta$-operators describe the low-lying eigenmodes of the transverse field Ising model and correspond to (gapped) pseudospin-wave excitations. The spectrum of these {\it auxiliary excitations} and their relation to physical properties is discussed in the next paragraph.  
\subsubsection{Pseudo-spin-wave mode and Mott-Hubbard gap}
\begin{figure}
\includegraphics[width=0.51\linewidth]{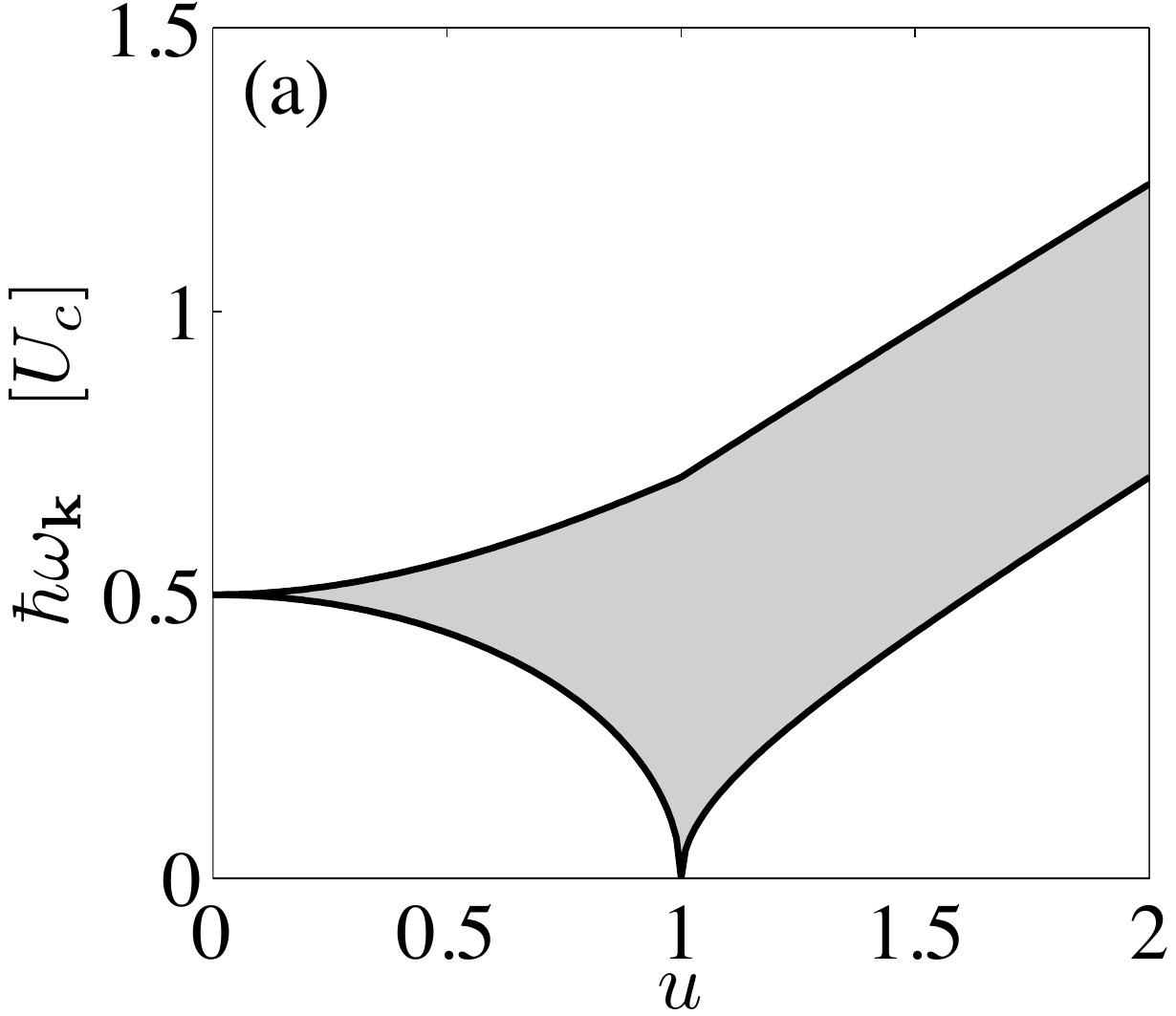}
\includegraphics[width=0.475\linewidth]{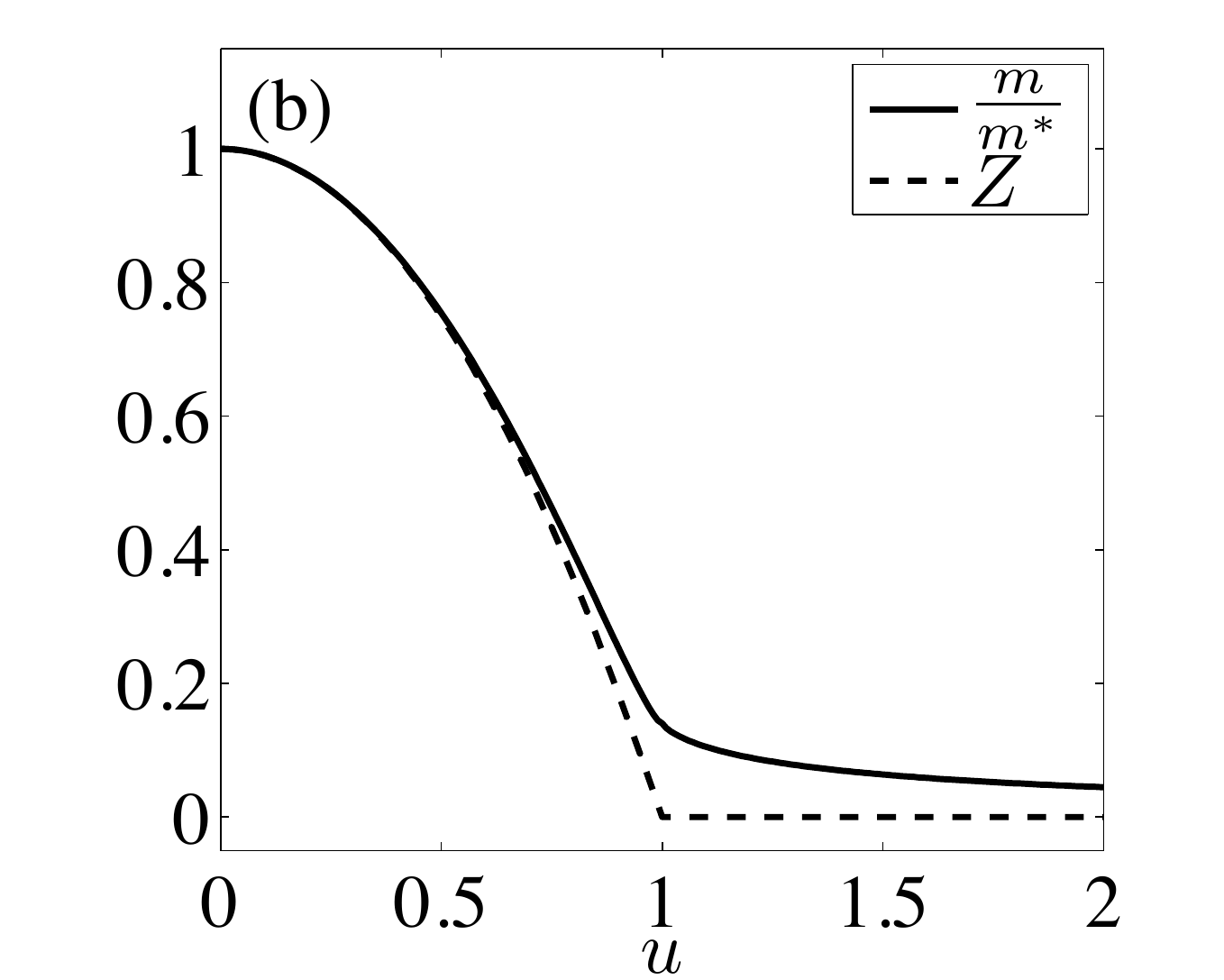}
\caption{(a) The band of pseudospin-wave excitations obtained in the spin-wave analysis of the transverse Ising model in three dimensions as function of $u=U/U_c$. (b) The inverse effective mass $m/m^*$ and the quasiparticle weight $Z$ as function of $U/U_c$. Note that at the Mott transition $u=1$, $Z$ vanishes in contrast to $m/m^*$.}
\label{fig:mode-mZ}
\end{figure}

From the diagonalization of the effective Hamiltonian we find the following pseudospin-wave spectrum 
\begin{equation}
\hbar\omega_{\bf k}=\frac{U_c}{2}\times\left\{
\begin{array}{cc}
\sqrt{1+u^2\gamma_{\bf k}},&{\rm for}\quad u\leq 1,\\
\sqrt{u^2+u\gamma_{\bf k}},&{\rm for}\quad u> 1,
\end{array}\right.
\label{eq:exmode}
\end{equation}
with an excitation gap $\Delta=\hbar\omega(0)$. The quantum criticality at $u=1$ is reflected in the softening of the mode (\ref{eq:exmode}). For $u>1$, the jump $\Delta\mu$ in the chemical potential from hole to particle doping amounts to twice the excitation gap,
\begin{equation}
\Delta\mu=2\Delta=U\sqrt{1-\frac{1}{u}}.
\label{eq:Deltamu}
\end{equation}
The above pseudospin-mode corresponds to the gapped charge excitation of the Mott insulator and Eq.~(\ref{eq:Deltamu}) coincides with the expression found for the Mott-Hubbard gap in the Kotliar-Ruckenstein formulation.\cite{Lavagna:1990,Castellani:1992} The band of the pseudospin mode (\ref{eq:exmode}) is shown in Fig.~\ref{fig:mode-mZ}(a) for different values of the interaction strength $u$.
\subsubsection{Renormalized ground state}
Quantum fluctuations lead to a renormalization of the ground-state energy $E_G^<$ ($u\leq 1$) and $E_G^>$ ($u>1$). We find
\begin{eqnarray*}
\frac{E_G^{<}}{N_s}&=&-\frac{U_c}{8}(1-u)^2-\frac{U_c}{8}\int_{\rm BZ}\!\!\frac{d^3k}{4\pi^3}\left(1-\sqrt{1+u^2\gamma_{\bf k}}\right),\nonumber \\
\frac{E_G^{>}}{N_s}&=&-\frac{U}{8}\int_{\rm BZ}\!\!\frac{d^3k}{4\pi^3}\left(1-\sqrt{1+\gamma_{\bf k}/u}\right),
\label{eq:renE}
\end{eqnarray*}
where the integrals over the Brillouin zone (BZ) represent quantum corrections to the result of the Gutzwiller approximation. Note that $E_G<0$ for any finite $u$, meaning that the trivial mean-field solution with $g=\chi=0$ is always higher in energy for any finite $u$. This is in contrast to the single-site mean-field approximation where the solution for $u>1$ coincides with the atomic limit. The present scheme based on the Schwinger boson representation of the pseudospin is analogous to taking into account Gaussian fluctuations around the mean-field transition without renormalizing the actual transition point. 

\subsubsection{Effective mass and quasiparticle weight in three dimensions}
As in the cluster mean-field treatment above and in previous studies invoking slave rotors,\cite{Florens:2004} we obtain that fluctuations reintroduce inter-site correlations beyond the mean-field value. In particular, there is a distinction between the $U$ dependence of the quasiparticle weight and of the effective mass renormalization of the quasiparticles, see Fig.~\ref{fig:mode-mZ}(b). While the hopping renormalization factor $g=m/m^*$ stays finite across the Mott transition, the quasiparticle weight $Z$ still vanishes for $u\rightarrow 1$. In the insulating phase, the effective mass is obtained from
\begin{eqnarray*}
g\equiv\frac{m}{m^*}&=&-\int\! d\varepsilon\, \rho_{\sigma}(\varepsilon)\frac{\varepsilon/D}{\sqrt{1+\frac{1}{u}\varepsilon/D}}\nonumber\\
&\approx&\frac{U_c}{2U}\int\! d\varepsilon\, \rho_{\sigma}(\varepsilon)\left(\varepsilon/D\right)^2=\frac{U_c}{12U},
\label{eq:mstarMI}
\end{eqnarray*}
where $D=6t$ is half of the band width and the second line holds in the limit $u\gg 1$. Thus, similar to the cluster mean-field calculation, the pseudospin-wave analysis captures the energy scale of the superexchange $J=4t^2/U$ for large $U$ and $m/m^*= J\chi/(8t)$ is finite due to virtual hopping processes which keep the number of doubly occupied sites small but finite. It is interesting to note that this value is the same as in the uniform resonating-valence-bond ($u$-RVB) phase of the Heisenberg model $H_S=J\sum_{\langle i,j\rangle}{\bf S}_i\cdot{\bf S}_j$ which is the strong-coupling limit of the low-energy sector of the Hubbard model at half filling. The $u$-RVB phase is obtained by a uniform spinon mean-field ansatz for a single real Hubbard-Stratonovich field $\chi$ [given by Eq.~\eqref{eq:chiij}]:\cite{Lee:2006b}
\begin{equation*}
J{\bf S}_i\cdot{\bf S}_j\rightarrow -\frac{J\chi}{8}\sum_{\sigma}\left(f_{i\sigma}^{\dag}f_{j\sigma}^{}+{\rm h.c.}\right),
\end{equation*}
which yields the same effective mass $\left(m/m^*\right)_{\rm u-RVB}=J\chi/(8t)$. The transition from a metal to a gapless spin liquid in frustrated geometries has recently been discussed by several authors.\cite{Morita:2002,Motrunich:2005,Lee:2005,Ohashi:2006,Ohashi:2008,Podolsky:2009,Yoshioka:2009}

\subsubsection{Spectral one-particle density}
We now turn to the discussion of how the auxiliary pseudspin excitations affect the physical excitation spectrum. To this end we calculate the spectral one-particle density $A_\sigma(\omega)$. We start by writing $A_\sigma(\omega)$ in the Lehmann representation
\begin{eqnarray*}
A_{\sigma}(\omega)&=&\sum_n\left[\left|\langle 0|c_{0\sigma}^{\phantom{\dag}}|n\rangle\right|^2\delta(\omega-\omega_{n0})\right.\nonumber\\
&&+\left.\left|\langle 0|c_{0\sigma}^{\dag}|n\rangle\right|^2\delta(\omega+\omega_{n0})\right].
\label{eq:Lehmann}
\end{eqnarray*}
where $|n\rangle$ denotes an eigenstate of the full Hamiltonian with energy $E_n$ and $\omega_{nm}=E_n-E_m$. In the slave-spin method used here, the true eigenstates are approximated by the mean-field eigenstates obtained in the pseudospin-wave analysis. We then have to calculate matrix elements of the form $\langle 0|I^x_0f_{{\bf q}\sigma}|n\rangle$. The details of the calculation are lengthy but straightforward. Here we present the main results.

The spectral weight contains a coherent quasiparticle sector $A_{\sigma}^{\rm coh}(\omega)$ as well as an incoherent contribution $A_{\sigma}^{\rm inc}(\omega)$. We find that the coherent contribution is given by
\begin{equation*}
A_{\sigma}^{\mathrm{coh}}(\omega)=\frac{Z}{g}\rho_{\sigma}(\omega/g).
\label{eq:Acoh}
\end{equation*}
Note that $A_{\sigma}(0)=A_{\sigma}^{\rm coh}(0)\propto Z/g$ gradually vanishes when approaching the Mott insulator, in contrast to the infinite dimensional result. In the metallic phase $u\leq 1$, the incoherent contribution is dominated by

\begin{figure}[b!]
\centering
\includegraphics[width=0.99\linewidth]{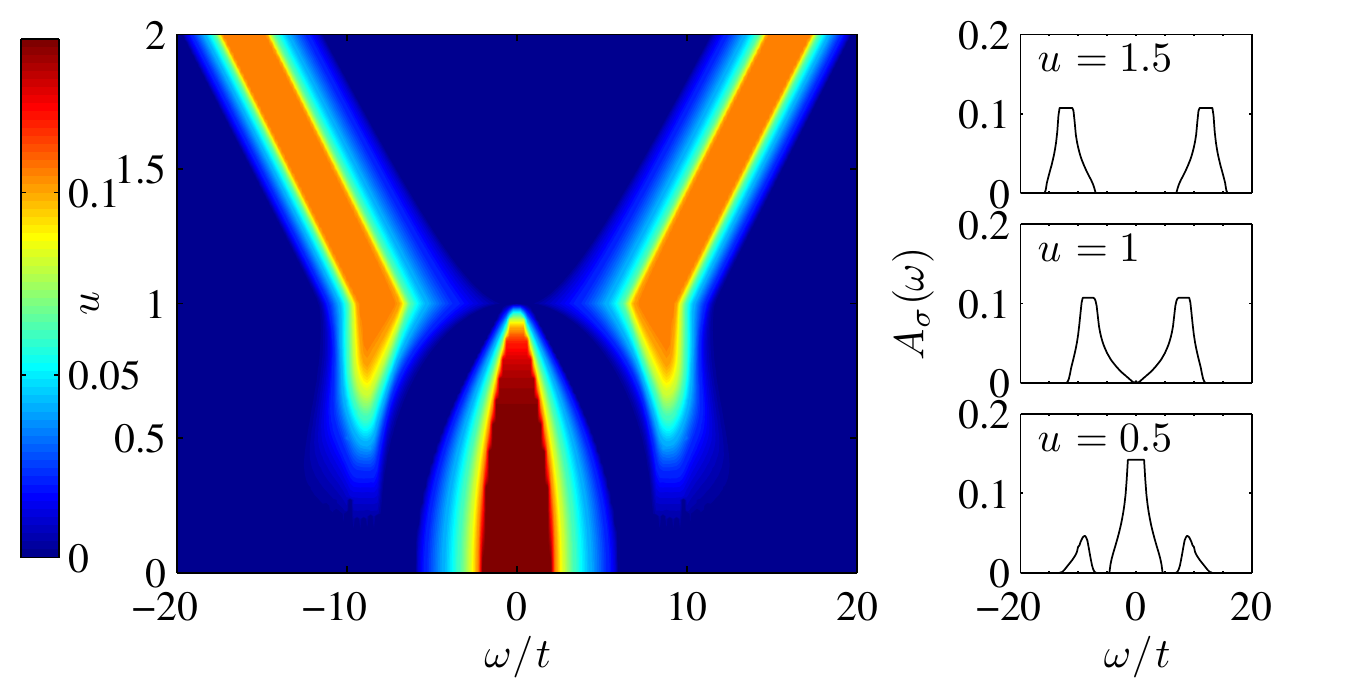}
\caption{(color online) The contour plot shows the one-particle spectral density $A_{\sigma}(\omega)$ with the coherent Gutzwiller and the (preformed) Hubbard bands as function of energy $\omega/t$ and interaction strength $u$. On the right-hand side, $A_{\sigma}(\omega)$ is shown for fixed values of $u$ in the metallic and the insulating phase.}
\label{fig:A1}
\end{figure}

\begin{widetext}
\begin{equation}
A_{\sigma}^{\mathrm{inc}}(\omega)\approx\frac{4D}{U_c}\int_0^{\infty}d\varepsilon\ \rho_{\sigma}(\varepsilon)\times\begin{cases}\rho_{\sigma}\left[\frac{D}{u^2}\left(\frac{4(\omega-g\varepsilon)^2}{U_c^2}-1\right)\right],& \Delta_-<\omega<\Delta_++gD;\\
\rho_{\sigma}\left[\frac{D}{u^2}\left(\frac{4(\omega+g\varepsilon)^2}{U_c^2}-1\right)\right],&-\Delta_+-gD<\omega<\Delta_-;\\
0,&\mathrm{else;}
\end{cases}
\label{eq:Ainc<}
\end{equation}
where $\Delta_{\pm}=\frac{U_c}{2}\sqrt{1\pm u^2}$ denote the edges of the excitation spectrum.
On the other hand, in the insulating phase $u>1$, the coherent part vanishes $A_{\sigma}^{\rm coh}(\omega)=0$ and we find
\begin{equation}
A_{\sigma}(\omega)=A_{\sigma}^{\mathrm{inc}}(\omega)=\frac{4D}{U_c}\int_0^{\infty}d\varepsilon\, \rho_{\sigma}(\varepsilon)\times\begin{cases}\rho_{\sigma}\left[\frac{D}{u}\left(\frac{4(\omega-g\varepsilon)^2}{U_c^2}-u^2\right)\right],& \Delta_-<\omega<\Delta_++gD;\\
\rho_{\sigma}\left[\frac{D}{u}\left(\frac{4(\omega+g\varepsilon)^2}{U_c^2}-u^2\right)\right],&-\Delta_+-gD<\omega<\Delta_-;\\
0,&\mathrm{else;}
\end{cases}
\label{eq:Ainc>}
\end{equation}
\end{widetext}
where $\Delta_{\pm}=\frac{U}{2}\sqrt{1\pm1/u}$. The spectral density is shown in Fig.~\ref{fig:A1} for different values of the interaction strength. The gapped mode (\ref{eq:exmode}) found in the transverse Ising model leads to the incoherent weight around $\pm\max(U_c,U)/2$ in the spectral density.\cite{Castellani:1992} In the metallic phase, we find the characteristic three peak structure with preformed Hubbard bands centered at $\hbar\omega\approx\pm U_c/2$ and a coherent Gutzwiller band at $\hbar\omega\approx 0$. The Gutzwiller band disappears at $u=1$ and the Hubbard bands touch at $\hbar\omega=0$. From the expression (\ref{eq:Ainc>}) we find that in the large $U$ limit the Hubbard bands assume a constant width of $W_{\rm HB}=U_c/2\approx 8t$ and are separated by $U$. This is consistent with the large $U$ expansion of the jump in the chemical potential (\ref{eq:Deltamu}) and the general expectation that both an added hole or an added double occupancy are mobile with an energy $-W_{\rm HB}/2$:
\begin{equation*}
\Delta\mu\approx U-\frac{U_c}{2}\equiv U-W_{\rm HB}.
\end{equation*}
The reduction in the bandwidth $W_{\rm HB}$ of the upper (or lower) Hubbard band as compared to the noninteracting bandwidth $W=2D=12t$ is in qualitative agreement with the retraceable-path approximation for a single hole doped into an infinite-$U$ Mott insulator.\cite{Brinkman:1970b}
\subsubsection{One-particle sum rule and fluctuation regime}
\begin{figure}
\centering
\includegraphics[width=0.7\linewidth]{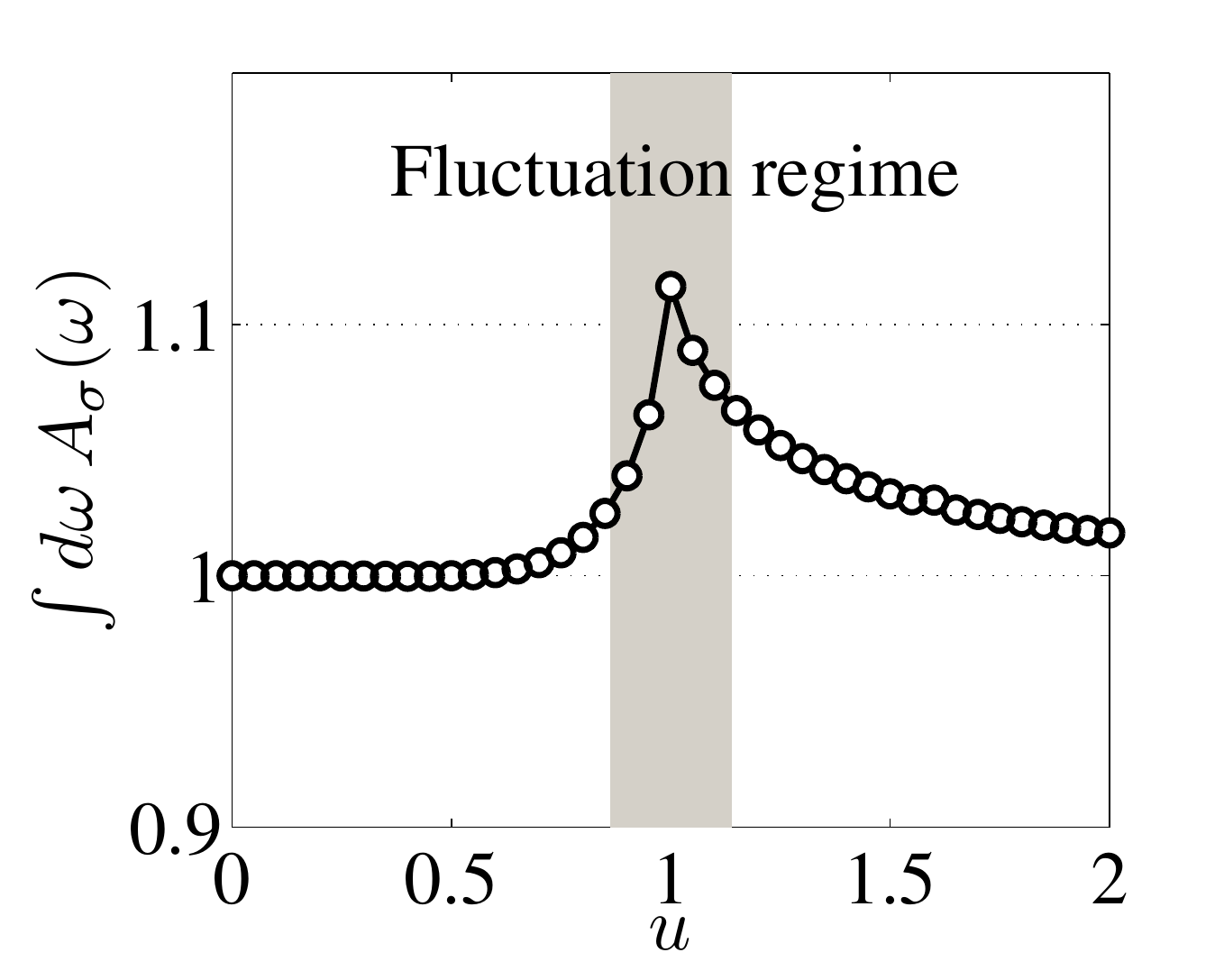}
\caption{The integrated spectral weight of the one-particle spectral density in the pseudospin-wave analysis. The shaded region denotes the fluctuation regime where the magnitude of the fluctuations is of the same order than the order parameter.}
\label{fig:spweight}
\end{figure}

To estimate the validity of the mean field plus spin-wave calculation, we compare the fluctuations with the magnitude of the order parameter and define $u^{\rm fl}$ through the condition 
\begin{equation*}
\frac{1}{1+6}\left[\sum_{\langle 0,i\rangle}\langle\delta I_0^x\delta I_i^x\rangle+\langle(\delta I_0^x)^2\rangle\right]=\langle I_0^x\rangle^2\bigg|_{u^{\rm fl}}.
\end{equation*}
For the cubic lattice considered here we obtain $u^{\rm fl}\approx 0.9$, hence, for $u\approx 1\pm 0.1$ the fluctuation induced corrections to the mean-field result are important and the validity of the above analysis is limited. 
In particular, the approximative nature of our treatment of the pseudospin problem violates $(I_i^x)^2=1/4$ and the spectral weight fails to be properly normalized. As shown in Fig.~\ref{fig:spweight}, the one-particle sum rule
\begin{equation*}
\int d\omega A_{\sigma}(\omega)=1,
\end{equation*}
is fulfilled within 10$\%$-12$\%$. The failure predominantly manifests itself in the fluctuation regime, however. Violations of sum rules are a known shortcoming of the Gaussian approximation. We expect that some of these inconsistencies can be cured by taking into account appropriate mode-mode couplings.

%
\section{Role of the constraint}
\label{sec:improving}
So far, we have discussed various aspects of the mean-field approximation where the eigenfunctions $|\Psi\rangle=|\Psi_I\rangle|\Psi_f\rangle$ of the slave-spin Hamiltonian \eqref{eq:Hprime} in the enlarged Hilbert space have been approximated by product states in pseudospin and pseudo fermion degrees of freedom and we have not paid attention on the relation \eqref{eq:conI}. There is an obvious shortcoming in this treatment: the expectation value $\langle I^z_i\rangle+1/2$ is in general not equal to $\langle\left(n_i-1\right)^2\rangle$ where $n_i=\sum_{\sigma}f_{i\sigma}^{\dag}f_{i\sigma}^{}$ is the local pseudo-fermion density. But since both operators are gauge invariant and the expectation values are equal for the physical state by construction we conclude that this relation should in fact hold for all eigenstates $|\Phi_m\rangle$ of $H'$:
\begin{equation}
\langle \Phi_m|I^z_i|\Phi_m\rangle+\frac{1}{2}=\langle\Phi_m|(n_i-1)^2|\Phi_m\rangle.
\label{eq:avcon}
\end{equation}
Therefore, we expect to obtain a better approximation of an eigenstate of $H'$ by enforcing the relation \eqref{eq:avcon} for mean-field states. This can be achieved by the usual Lagrange multiplier method. In this way, we can access a larger class of correlations in the mean-field approximation and a prominent example missed so far is magnetism. In the following we discuss two possibilities to satisfy \eqref{eq:avcon} for mean-field states. The first approach relies on the presence of a finite staggered magnetization $\nu_s$ and is therefore expected to be of relevance for magnetically unfrustrated models. The second approach involves an iteration of the slave-spin approximation in the paramagnetic sector and its natural applications are therefore models where magnetic order is fully frustrated. In this brief overview, we do not analyze weakly frustrated systems such as the $t-t'$ model discussed in Sec.~\ref{subsec:FS}. For these models the situation is more complex since aspects such as the Fermi surface deformation are likely to play a role as well.

We follow a standard procedure to enforce the relation \eqref{eq:avcon} for mean-field states by adding a term
\begin{equation}
\Lambda=-\frac{1}{2}\sum_i\lambda_i\left[I^z_i+\frac{1}{2}-\left(n_i-1\right)^2\right],
\label{eq:lagrange}
\end{equation} 
to the slave-spin Hamiltonian \eqref{eq:Hprime}. One then seeks for a stationary solution with respect to the Lagrange multiplier $\lambda_i$:
\begin{equation}
\frac{\partial \langle H'+\Lambda\rangle}{\partial\lambda_i}=0.
\label{eq:Hlam}
\end{equation}
\subsection{N\'eel state in unfrustrated systems}
On a bipartite lattice a N\'eel-ordered state is usually a leading instability and in the following we explore the possibility that the term \eqref{eq:lagrange} triggers this instability. Furthermore, introducing an order parameter for the staggered magnetization $\nu_s$ allows to satisfy the constraint \eqref{eq:conI} on average, see also Eq.~\eqref{eq:avcon}. Hence, we decouple \eqref{eq:lagrange} in the antiferromagnetic channel by writing (at half filling)
\begin{equation}
n_{i\uparrow}n_{i\downarrow}\rightarrow-\frac{\nu_i}{2}(n_{i\uparrow}-n_{i\downarrow})+\frac{1}{2}n_i-\frac{1-\nu_i^2}{4},
\label{eq:decoupling}
\end{equation}
where $\nu_i=\langle n_{i\uparrow}-n_{i\downarrow}\rangle$ and we assume $\nu_i=\nu_s$ if $i$ belongs to the $A$ sublattice and $\nu_i=-\nu_s$ for $i$ on the $B$ sublattice. Furthermore, we seek for a translational invariant state with $\lambda_i=\lambda$. In addition to the slave-spin self-consistency \eqref{eq:gij} and \eqref{eq:chiij} there are two more self-consistency equations in order to determine the parameters $g$, $\chi$, $\lambda$ and $\nu_s$:
\begin{eqnarray}
g&=&4\langle I_i^xI_j^x\rangle,\nonumber\\
\chi&=&4\sum_{\sigma}\left(\langle f_{i\sigma}^{\dag}f_{j\sigma}\rangle_f+{\rm c.c}\right),\nonumber\\
\frac{\nu_s^2}{2}&=&-\langle I_i^z\rangle,\label{eq:Connu}\\
\frac{1}{\lambda}&=&2\int_{-zt}^{\varepsilon_F=0}d\varepsilon\frac{\rho_{\sigma}(\varepsilon)}{\sqrt{4g^2\varepsilon^2+\lambda^2\nu_s^2}}\label{eq:SCgap}.
\end{eqnarray}
Equation~\eqref{eq:Connu} is just the average constraint using the decoupling \eqref{eq:decoupling} and \eqref{eq:SCgap} is the gap equation for the pseudo-fermion problem. 

\begin{figure}
\centering
\includegraphics[width=0.8\linewidth]{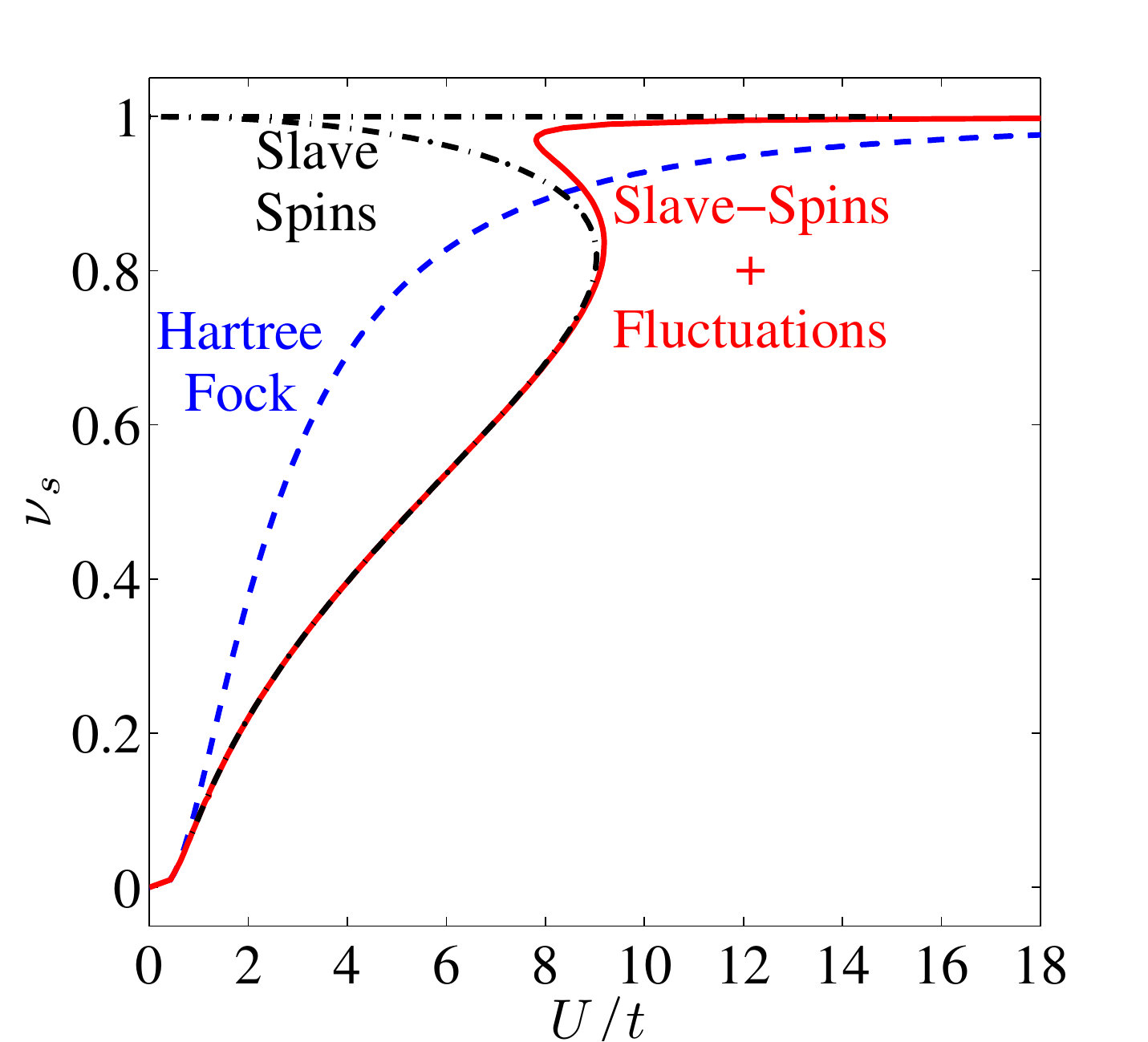}
\caption{(color online) The staggered magnetization $\nu_s$ at $T=0$ for the square lattice as function of the interaction strength $U/t$ calculated using the local slave-spin approximation of Sec.~\ref{subsec:local}(black, dashed-dotted) and with fluctuation corrections as discussed in Sec.~\ref{sec:3DMott} (red, full). For comparison we also show the result obtained in the Hartree-Fock mean-field theory (blue, dashed). In the slave-spin calculation there is a finite range of $U/t$ where there exists two solutions for $\nu_s$ which corresponds to a (local) minima of the energy.}
\label{fig:nus}
\end{figure}

We have solved these equations for different values of $U/t$ on a square lattice with nearest-neighbor hopping $t$. The result for $\nu_s$ is summarized in Fig.~\ref{fig:nus} for the local slave-spin approximation (Sec.~\ref{subsec:local}) and with fluctuations (Sec.~\ref{sec:3DMott}). Also shown is the result obtained in the standard Hartree-Fock (HF) mean-field theory applied to the original model \eqref{eq:Hhubb}. We find a first-order transition in the staggered magnetization both in the local approximation and with the inclusion of fluctuations in the pseudospin-wave sector. Moreover, for a finite range of $U/t$, the self-consistency equations have two distinct solutions for $\nu_s$ which both locally minimize the energy. One solution corresponds to a ``low-$\nu_s$'' phase of delocalized character (``Slater") the other to a ``high-$\nu_s$'' phase of localized character (``Mott-Heisenberg"). (However, the naive decoupling \eqref{eq:decoupling} should be taken with care, in particular for larger values of $U/t$.)

As seen in Fig.~\ref{fig:nus}, the present approach coincides with the pure HF of the original model only in lowest order in $U/t$, but clearly differs in higher orders. These pronounced differences are attributed to a ``positive feedback effect" between the staggered magnetization $\nu_s$ and the hopping renormalization factor $g$ which, formally, enters by the averaged constraint \eqref{eq:Connu}. In the low-$\nu_s$ phase, the system optimizes the kinetic energy by sustaining a weak staggered magnetization but a small effective mass (large $g$). $\nu_s$ is therefore smaller than in HF.  On the other hand, the high-$\nu_s$ solution results from the optimization of the potential energy: the physical mechanism behind it is the fact that an enhanced effective mass (localization) favors the formation of local magnetic moments (singly occupied sites) which allows to minimize the potential energy cost. $\nu_s$ is therefore larger compared with the HF result. 

Including fluctuations in the slave-spin problem reduces the regime where the high-$\nu_s$ solution is stable. It is likely that the quantum fluctuations in $\nu_s$, which have not been considered here, will reduce the stability regime even more and it is an open problem if the first-order transition survives once we go beyond the mean-field decoupling \eqref{eq:decoupling}. (We note that in dynamical mean-field theory on the infinite dimensional hypercubic lattice a crossover is observed.)\cite{Pruschke:2003} Furthermore, such fluctuations lead to an overall reduction in the staggered moment also in the large $U/t$ limit. From spin-wave theory\cite{Huse:1988} and Monte Carlo simulations\cite{Reger:1988} of the spin-1/2 Heisenberg model on the square lattice it is known that the sublattice magnetization is only around 60\% of its classical value.

\subsection{Iterated slave-spin scheme for full frustration}
We now discuss an alternative way to satisfy the constraint on average which does not rely on the presence of a magnetic order parameter. This scheme is therefore more likely to play a role in fully frustrated systems where magnetic order is suppressed. Also in this approach we find a first-order transition between an itinerant and a localized phase as function of $U/t$. 

The term \eqref{eq:lagrange} introduces an onsite interaction of strength $\lambda$ (we assume $\lambda_i=\lambda$) for the $f$ fermions in the mean-field treatment and $H_f$ takes the form of a Hubbard model with renormalized parameters: effective hopping $g_It_{ij}$ and onsite interaction $\lambda$. Let us again use the slave-spin mean-field theory to treat this renormalized model: following Sec.~\ref{sec:slavespinformulation} we introduce an additional set of pseudofermions $h_i^{(\dag)}$ and pseudopins ${\bf K}_i$ to represent the $f$ fermions and the eigenstates of $H_f$ are approximated as product states $|\Psi_f\rangle=|\Psi_K\rangle|\Psi_h\rangle$. In order to satisfy the original constraint between the $f$ fermions and the ${\bf I}$ spins on average, we can relax the constraint between the $h$ fermions and the ${\bf K}$ spins. The reason is that the relation \eqref{eq:IvsD} for the $(h,{\bf K})$ pair gives 
\begin{equation*}
\Big\langle \Big(\sum_{\sigma} f_{i\sigma}^{\dag}f_{i\sigma}^{}-1\Big)^2\Big\rangle\equiv\langle K_i^z\rangle+\frac{1}{2}.
\end{equation*}
But from \eqref{eq:Hlam} it follows that $\langle K_i^z\rangle=\langle I_i^z\rangle$ and the above relation is just the average constraint for the $(f,{\bf I})$ pair. Equation~\eqref{eq:Hlam} also implies that the ratio of the transverse field to the exchange coupling is the same for the ${\bf I}$  and the ${\bf K}$ model and that $g_I=g_K\equiv g$. 
As a result, we find that $\lambda=U/2$. In the local approximation for the transverse field Ising model, a non-trivial solution exists only for $v\leq1$ where $v=4u/(3\sqrt{3})$. The kinetic energy is renormalized by $g^2$ where for $v\leq1$
\begin{equation*}
g=\frac{1+2\cos(\phi/3)}{3},
\end{equation*}
with $\phi=\arccos(1-2v^2)$ and $g=0$ otherwise. We thus find a first order transition at a reduced critical interaction strength $4U_c/(3\sqrt{3})$ between a paramagnetic metal and an insulator.

We note here that a paramagnetic first-order transition at $T=0$ has recently been reported\cite{Yoshioka:2009} for the single-band Hubbard model on the (highly-frustrated) triangular lattice. These calculations revealed a first-order transition from a paramagnetic metal to a paramagnetic insulator. Moreover, a second first-order transition at larger values of $U/t$ between the paramagnetic metal and the $120^{\circ}$ N\'eel-ordered state was found.

\section{Conclusions}
We have reviewed a slave-particle formulation for strongly interacting Fermi systems which can be considered as a minimal formulation of previous representations. Our approach is based on auxiliary pseudospin variables displaying a local $\mathsf{Z}_2$ gauge freedom. They refer to the local charge modulo two. The simplicity of the representation allows to exactly solve the non-interacting model and to gain insights into the artificial local symmetry introduced by enlarging the Hilbert space. For the interacting model, the main focus so far has been the investigation of the corresponding mean-field approximation. We have proposed an interpretation of the mean-field decoupling in terms of approximating a ``representative eigenstate" in the enlarged Hilbert space rather than the physical eigenstate. In this light, our mean-field results do not contradict Elitzur's theorem and we have argued that reasonable results for gauge invariant observables can be obtained.

The mean-field theory has been applied to various aspects related to the interaction-driven Mott transition in the single-band Hubbard model. The role of inter-site correlations present in finite dimensions has been discussed in a cluster approximation and we have found that inter-site correlations deform the Fermi surface toward the fully nested one. By including quantum fluctuations in the auxiliary Ising model, we have analytically calculated the single-particle density of state and have elaborated the connection between the (gapped) pseudospin-wave excitations and the (preformed) Hubbard bands of the original model.

We have also commented on the importance to enforce an averaged constraint which has to be satisfied by all eigenstates in the enlarged Hilbert space. In this way it is also possible to include magnetic correlations. For the nearest-neighbor hopping model on the square lattice we have found a first order transition between a phase with a low and high value of the staggered magnetization.

There are many more open problems left for future studies and we hope to stimulate some further investigations. For example, an extension of the present approach away from half-filling is highly desirable; in particular in view of potential applications of the current formalism to inhomogeneous systems (such as artificially structured correlated materials\cite{Rueegg:2007,Rueegg:2008a} or cold atomic gases in optical traps). Furthermore, we encourage an investigation of possible ``flux patterns" (and its excitations) associated with the local $\mathsf{Z}_2$ gauge freedom on the mean-field level. In order to go beyond the mean-field description it is an interesting problem to study the $\mathsf{Z}_2$ lattice gauge theory coupled to the ``matter field" of the pseudo fermions as obtained from the present approach. 
Moreover, such a study could also help to make contact with a previously considered $\mathsf{Z}_2$ gauge theory for strongly correlated electrons\cite{Senthil:2000} and could add much to our understanding of the present formulation and its connection to various other slave-particle representations.

\begin{acknowledgements}
We thank E. Altman, G. Blatter, G.~A. Fiete, F.~Hassler, D. Podolsky, O.~I.~Motrunich, and T.~M. Rice for stimulating discussions. S.D.H acknowledges financial support by the Center for Theoretical Studies at ETH Z\"urich. A.R. acknowledges support from ARO Grant No. W911NF-09-1-0527. This study was financially supported by the NCCR MaNEP of the Swiss Nationalfonds.
\end{acknowledgements}

\begin{appendix}
\section{Relation to the four boson and the two-spin formulation}
\label{app:relation}
In this appendix we relate our approach to the KR {\it four} boson formulation for a one-band model.\cite{Kotliar:1986} An equivalent representation is also obtained by using {\it two} slave spins introduced by de'Medici and co-workers in Ref.~\onlinecite{deMedici:2005}. We show that our representation is obtained by restricting the action of the two aforementioned slave spins to the particle-hole and spin-$\sigma$ symmetric subspace. 

In the following, it is sufficient to focus on the local Hilbert space. Kotliar and Ruckenstein introduced four Bose creation and annihilation operators $e^{(\dag)}$, $p_{\sigma}^{(\dag)}$ and $d^{(\dag)}$ as well as two Fermi operators $f_{\sigma}^{(\dag)}$ to represent the local occupation number states as
\begin{equation*}
|e\rangle=e^{\dag}|\Omega\rangle,\quad |\sigma\rangle=p_{\sigma}^{\dag}f_{\sigma}^{\dag}|\Omega\rangle, \quad |d\rangle=d^{\dag}f_{\uparrow}^{\dag}f_{\downarrow}^{\dag}|\Omega\rangle,
\label{eq:KR}
\end{equation*}
where $|\Omega\rangle$ is a fake vacuum state. The constraints to project out unphysical states are given by
\begin{eqnarray}
e^{\dag}e+\sum_{\sigma}p_{\sigma}^{\dag}p_{\sigma}^{}+d^{\dag}d&=&1,
\label{eq:KRcon1}\\
d^{\dag}d+p_{\sigma}^{\dag}p_{\sigma}^{}&=&f_{\sigma}^{\dag}f_{\sigma}^{}.\label{eq:KRcon2}
\end{eqnarray}
The above constraints also imply that 
\begin{equation}
d^{\dag}d=f_{\uparrow}^{\dag}f_{\uparrow}f_{\downarrow}^{\dag}f_{\downarrow},
\label{eq:conD}
\end{equation}
which is analog to Eq.~\eqref{eq:conI}. The physical annihilation and creation operators can be represented as
\begin{equation}
c_{\sigma}=z_{\sigma}f_{\sigma},\quad c_{\sigma}^{\dag}=z_{\sigma}^{\dag}f_{\sigma}^{\dag},
\label{eq:c}
\end{equation}
where the simplest form for the $z$-operators is
\begin{equation*}
z_{\sigma}=p_{\bar{\sigma}}^{\dag}d+e^{\dag}p_{\sigma},\quad z_{\sigma}^{\dag}=d^{\dag}p_{\bar{\sigma}}+p_{\sigma}^{\dag}e.
\end{equation*}
(In order to reproduce the correct non-interacting limit in the mean-field theory, a modified form was introduced.)\cite{Kotliar:1986} The constraint (\ref{eq:KRcon1}) can be resolved by introducing slave-spin variables \cite{deMedici:2005} according to 
\begin{equation*}
S_{\sigma}^+= z_{\sigma}^{\dag},\quad S_{\sigma}^-= z_{\sigma},\quad S_{\sigma}^z+\frac{1}{2}= d^{\dag}d+p_{\sigma}^{\dag}p_{\sigma}^{}.
\end{equation*}
The constraint \eqref{eq:KRcon2} then reads 
\begin{equation}
S_{\sigma}^z+\frac{1}{2}=f_{\sigma}^{\dag}f_{\sigma}^{}.
\end{equation}
We note that the slave-spin representation of Ref.~\onlinecite{deMedici:2005} differs from the one we have used in this paper since $S_{\sigma}^z+1/2$ refers to the presence $(S_{\sigma}^z=+1/2)$ or absence $(S_{\sigma}^z=-1/2)$ of a spin-$\sigma$ electron. In contrast, in our method, only one slave-spin ${\bf I}$ is introduced and the $I^z$-value refers to the {\it total local charge modulo two.} In the two slave-spin method, the physical annihilation operator was represented as 
\begin{equation}
c_{\sigma}=\left[S_{\sigma}^-+\alpha(n) S_{\sigma}^{+}\right]f_{\sigma},
\end{equation}
which in the physical subspace is equivalent to Eq.~(\ref{eq:c}). The parameter
\begin{equation*}
\alpha(n)=\frac{2}{\sqrt{n(2-n)}}-1
\end{equation*}
was adjusted such that the correct $U=0$ limit at particle density $n$ is recovered.\cite{Hassan:2010} In particular, for $n=1$, the physical annihilation operator was written in the form
\begin{equation}
c_{\sigma}=2S_{\sigma}^x f_{\sigma}.
\end{equation}
Since $(2S_{\sigma}^x)^2=1$ it follows that for $n=1$ the canonical anti-commutation relation is preserved in the mean-field decoupling, similar to Eq.~(\ref{eq:anti}). However, for $n\neq1$, this property is lost. It seems that it is generally difficult to find a representation which preserves the anticommutation relations {\it and} reproduces the correct noninteracting limit away from half-filling.

The connection to our reduced slave-spin representation at $n=1$ is achieved by restricting the slave-spin operators to the two-dimensional particle-hole and spin-$\sigma$ symmetric subspace spanned by
\begin{equation*}
|-\rangle=|0,1\rangle, \quad|+\rangle=\frac{1}{\sqrt{2}}\left(|1,1\rangle+|-1,1\rangle\right).
\end{equation*}
Here, we have introduced the triplet states $|m_z,1\rangle$ ($m_z=0,\pm1$) of the total spin $\vec{S}=\vec{S}_{\uparrow}+\vec{S}_{\downarrow}$. The states $|\pm\rangle$ can be viewed as the eigenstates of a pseudospin operator $I^z$, see Eq.~\eqref{eq:pm}. This establishes the connection between the different representations.
\end{appendix}

\bibliography{referencesPhD}

\begin{thebibliography}{70}
\bibitem[{not({\natexlab{a}})}]{note:P}
\bibinfo{note}{Given any eigenstate $|\Phi_n\rangle$ of $H'$ which does not
  exclusively lie in the unphysical subspace, the (unnormalized) physical
  eigenstate is obtained by projection:
  $|\Psi_n\rangle=\mathcal{P}|\Phi_n\rangle$.}
\end{thebibliography}
\end{document}